\documentclass{aa}
\usepackage{txfonts}
\usepackage{natbib}
\usepackage{graphicx,amssymb}
\usepackage{color}
\usepackage{longtable}
\usepackage{multirow}
\newcommand{\wave}[1]{$\lambda#1\,\mathrm{cm}$}  % one wavelength
\newcommand{\HI}{\mathrm{H\,\scriptstyle I}}
\newcommand{\HII}{\mathrm{H\,\scriptstyle II}}

\bibpunct{(}{)}{;}{a}{}{,} % to follow the A&A style
\def\alphan{\alpha_{\rm n}}
\def\cm{\,{\rm cm}}

\def\degr{\hbox{$^\circ$}}
\def\farcm{\hbox{$.\mkern-4mu^\prime$}}

\def\radm{\,\mathrm{rad\,m^{-2}}}
\begin{document}

   \title{Magnetic fields and cosmic rays in M~31
   \thanks{Based on observations with the 100-m telescope of the Max-Planck-Institut f\"ur Radioastronomie at Effelsberg.}}
   \subtitle{I. Spectral indices, scale lengths, Faraday rotation, and magnetic field pattern}

\titlerunning{Magnetic fields and cosmic rays in M~31}
\authorrunning{R.~Beck et al.}

   \author{R. Beck
          \and
          \& E.M. Berkhuijsen
          \and R. Gie{\ss}{\"u}bel
          \thanks{Present e-mail address: r.giessuebel@gmx.de}
          \and D.D. Mulcahy
          \thanks{Present e-mail address: david.dec.mulcahy@gmail.com}
          }

   \institute{Max-Planck-Institut f\"ur Radioastronomie, Auf dem H\"ugel 69, 53121 Bonn, Germany\\
              \email{rbeck@mpifr-bonn.mpg.de}
             }

   \date{Received 8 August 2019; accepted 16 October 2019}

% \abstract{}{}{}{}{}
% 5 {} token are mandatory

  \abstract
      {Magnetic fields play an important role in the dynamics and evolution
        of galaxies; however, the amplification and ordering of the initial
        seed fields are not fully understood. The nearby spiral galaxy
        M~31 is an ideal laboratory for extensive studies of magnetic fields.}
      {Our aim was to measure the intrinsic structure of the magnetic fields in
        M~31 and compare them with dynamo models of field amplification.}
      {The intensity of polarized synchrotron emission and its orientation
        are used to measure the orientations of
        the magnetic field components in the plane of the sky. The Faraday
        rotation measure gives information about the field components
        along the line of sight. With the Effelsberg 100-m telescope three
        deep radio continuum surveys of the Andromeda galaxy, M~31, were
        performed at 2.645, 4.85, and 8.35\,GHz (wavelengths of 11.3, 6.2, and 3.6\,cm). The \wave{3.6} survey is the first radio survey of M~31
        at such small wavelengths. Maps of the Faraday rotation measures
        ($RMs$) are calculated from the distributions of the polarization angle.}
      {At all wavelengths the total and polarized emission is concentrated
        in a ring-like structure of about 7--13\,kpc in radius
        from the centre. Propagation of cosmic rays away from the star-forming
        regions is evident. The ring of synchrotron emission is wider than
        the ring of the thermal radio emission, and the radial scale length
        of synchrotron emission is larger than that of thermal emission.
        The polarized intensity from the ring in the plane of the sky varies
        double-periodically with azimuthal angle, indicating that the
        ordered magnetic field is oriented almost along the ring, with a
        pitch angle of $-14\degr \pm 2\degr$ at \wave{6.2}.
%   The diffusion length along the ring, parallel to the disc, is $\simeq$1.5\,kpc and the diffusion coefficient is
%   $D_\mathrm{E}\simeq 2 \cdot 10^{28}$\,cm$^2$\,s$^{-1}$. Since M~31 has only a thin radio disc, $D_\mathrm{E}$ perpendicular to the disc is
%   smaller by about a factor of 10.
        The $RM$ varies systematically along the ring. The analysis shows a
        large-scale sinusoidal variation with azimuthal angle,
        signature of an axisymmetric spiral ($ASS$) regular magnetic field,
        plus a superimposed double-periodic variation of a bisymmetric
        spiral ($BSS$) regular field with about six times smaller amplitude.
        %The $RM$ offset of $-125\pm2$\,rad/m$^2$ originates in the Galactic foreground.
        The $RM$ amplitude of $(118\pm3)\,\radm$ between \wave{6.2} and \wave{3.6}
        is about 50\% larger than between \wave{11.3} and \wave{6.2}, indicating
        that Faraday depolarization at \wave{11.3} is stronger
        (i.e. with a larger Faraday thickness) than at \wave{6.2} and \wave{3.6}.
        The phase of the sinusoidal $RM$ variation of $-7^\circ \pm 1^\circ$
        is interpreted as the average spiral pitch angle of the regular field.
        The average pitch angle of the ordered field, as derived from the intrinsic
        orientation of the polarized emission (corrected for Faraday rotation), is
        significantly smaller: $-26\degr \pm 3\degr$.
        }
      {The dominating axisymmetric (ASS) plus the weaker bisymmetric (BSS)
        field of M~31 is the most compelling case so far of a field generated
      by the action of a mean-field dynamo. The difference in pitch
      angle of the regular and the ordered fields indicates that the ordered field
      contains a significant fraction of an anisotropic turbulent field that has a
      different pattern than the regular ($ASS + BSS$) magnetic field.
      }


   \maketitle
%
%---------------------SECTION------1---------------------------------------

\section{Introduction}
\label{sec:intro}

%---------------------------------------------------------------------------
%TABLE1
\begin{table}
\begin{center}
\caption{Basic parameters of M~31 $\equiv$ NGC~224.}
\label{tab:parameter}
\begin{tabular}{lc}
\hline
RA [J2000] 							& 00h 42m 46.1s \\
DEC [J2000]							& $+41\degr$ $16\arcmin$ $12\arcsec$ \\
Distance [kpc]					    & $780\pm40$ $^1$  \\
Inclination [$^\circ$] 				& 75 $^2$   \\
PA [$^\circ$]						& 37 $^2$   \\
v$_{\text{sys}}$ [km s$^{-1}$] 	& $-305\pm7$ $^3$		\\
v$_{\text{rot}}$ [km s$^{-1}$] (disc) 	& 230--275 $^3$	    \\
Approaching side 			& south-west	\\
Receding side 			  	& north-east	\\
SFR [M$_{\odot}$~yr$^{-1}$]	&	0.3 $^4$ 		\\		
Classification 			  	& SAS3 $^5$	    \\
\hline
\end{tabular}
%\begin{tablenotes}
\footnotesize
\item References:
\item[1] \citet{stanek98}
\item[2] \citet{berkhuijsen77,chemin09}
\item[3] \citet{chemin09}
\item[4] \citet{taba10,rahmani16}
\item[5] \citet{devaucouleurs76}
%\end{tablenotes}
\end{center}
\end{table}
\normalsize
%---------------------------------------------------------------------------

%---------------------------------------------------------------------------
%TABLE2
\begin{table*}
\begin{center}
  \caption{New radio continuum surveys conducted with the Effelsberg 100-m telescope.}
  \label{tab:surveys}
  \begin{tabular}{cccccccc}
  \hline
  Central freq. & Bandwidth & Wavelength & HPBW $^1$ & Map size & rms noise ($I$) $^2$ & rms noise ($PI$) $^3$ & Obs. dates \\
 $[$GHz$]$ & $[$MHz$]$ & [cm] & [$\arcmin$] & [$\arcmin$] & [mJy/beam] & [mJy/beam] &  \\
  \hline
  2.645 &  70 & 11.33 & 4.4 & 196 x 92 & 1.2 & 0.4 & Oct 2010 \\
  4.85 &  300 &  6.18 & 2.6 & 140 x 80 & 0.3 & 0.05 &  June 2001--Aug 2005 \\
  8.35 & 1100 &  3.59 & 1.4 & 116 x 40 & 0.25--0.3 $^4$ & 0.06--0.12 $^4$& Dec 2001--Sept 2012 \\
  \hline
  \end{tabular}
\footnotesize
\item Notes:
\item[1] Half-power beamwidth
\item[2] Noise in total power ($I$) has an uncertainty of about 20--30\% because of
  residual scanning effects.
  Noise in $I$ is\\
  generally larger than that in polarized intensity ($PI$) due to confusion by
  many weak unpolarized background sources.
  \item[3] Noise in $PI$ has an uncertainty of about 10\%. Noise in Stokes
    $Q$ and $U$ is very similar to that in $PI$.
  \item[4] The lower value refers to the noise in the inner part of the map
    ($40\arcmin \times 40\arcmin$).
\end{center}
\end{table*}
\normalsize
%---------------------------------------------------------------------------

Interstellar magnetic fields play an important role in the structure
and evolution of galaxies. They provide support to the gas against the
gravitational field \citep{boulares90}, affect the star formation rate
\citep{krumholz19} and the multiphase structure of the ISM \citep{evirgen19},
regulate galactic outflows and winds \citep{evirgen19}, and
control the propagation of cosmic rays \citep[e.g.][]{zweibel13}.

Magnetic fields can be turbulent, ordered, or regular, and are generated
by different physical processes. Turbulent fields are amplified by turbulent gas
motions, called the small-scale dynamo \citep[e.g.][]{brandenburg05}.
%Turbulent gas motions can also tangle ordered fields.
Ordered fields, obtained from turbulent fields by compressing and shearing
gas flows, reverse their sign on small scales and are called anisotropic turbulent
fields. On the other hand, ordered fields generated by the
mean-field $\alpha$--$\Omega$ dynamo \citep{ruzmaikin88,beck96,chamandy16}
reveal a coherent direction over several kpc and are called regular fields or
mean fields \citep[for a review see][]{beck15c}.

Synchrotron radio emission is the best tool for studying
magnetic fields in 3D without effects due to absorption.
Synchrotron intensity is a measure of the strength of the field
components on the sky plane and the density of cosmic-ray electrons.
Linearly polarized synchrotron emission is a signature of the strength
and orientation of ordered fields in the plane of the sky.
Faraday rotation of the
polarization angle increases with the square of the wavelength, the density of
thermal electrons, and the strength of regular fields along the line of
sight; the sign of the Faraday rotation gives the field direction.
Unpolarized synchrotron emission traces turbulent fields (or ordered
fields tangled by turbulent gas motions) that cannot be resolved by the
telescope beam. Turbulent or tangled fields are strongest
in star-forming regions in the gaseous arms of spiral galaxies
\citep{beck07,taba13a} with energy densities similar to that of
the turbulent kinetic energy of the gas \citep{beck15c}. Ordered fields
reveal spiral patterns in most galaxies observed so far.
Measuring the azimuthal variation of Faraday rotation can reveal
large-scale modes of the regular field that are generated by the mean-field
dynamo \citep{krause90}.

The Andromeda galaxy, M~31, is particularly suited to investigating
interstellar magnetic fields thanks to its proximity and prominent `ring'
of star formation.
%M~31 is the prototype of a galaxy hosting a magnetic field
%generated by the mean-field $\alpha$--$\Omega$ dynamo \citep{beck96}.
The radio emission and magnetic field properties of M~31
have been studied extensively with the Effelsberg 100-m,
the Very Large Array (VLA), and
the Westerbork Synthesis Radio Telescope (WSRT)
\citep{berkhuijsen77,berkhuijsen83,beck80,beck82,beck89,beck98,berkhuijsen03,giessuebel13,giessuebel14}.
The total and polarized emissions are concentrated in a ring-like structure with a radius of
about 7--13\,kpc   from the centre, the region with the
highest density of cold molecular gas \citep{nieten06}, warm neutral gas
\citep{brinks84,braun09,chemin09}, warm ionized gas \citep{devereux94},
and dust \citep{gordon06,fritz12}; it is the main location of present-day
star formation \citep[e.g.][and references therein]{taba10,rahmani16}.

The first radio polarization survey of M~31 was performed with the
Effelsberg telescope at 2.7\,GHz (\wave{11.1}) \citep{beck80}. Faraday rotation
measures were estimated from the differences between the orientations of the observed
polarization vectors and those of a regular field with a constant direction along
the azimuthal and radial directions in the emission ring \citep{sofue81,beck82}.
The angle differences showed a clear sinusoidal
variation with azimuthal angle. This result, as well as the double-periodic azimuthal
variation of polarized intensity, were found to be consistent with the ring-like field
\citep{beck82}. The phase shift of the angle differences relative to the major axis
indicated that the field pattern is not a ring, but a tightly wound spiral with a pitch
angle of about $-10\degr$ \citep{ruzmaikin90}.\footnote{A negative pitch angle indicates
that the spiral pattern is trailing with respect to the global rotation.}
%a trailing spiral pattern

These authors interpreted this regular axisymmetric field with a spiral pattern
($ASS$) as the lowest mode excited by a large-scale ($\alpha$--$\Omega$) dynamo.
Indication of a superposition of the next higher mode with a lower
amplitude, the bisymmetric spiral field ($BSS$), was found by
\citet{sofue87a} and \citet{ruzmaikin90}. However, the above results were
based on observations at one single frequency, and thus had to rely on
the assumption of a simple field geometry.

Completion of a polarization survey of M~31 at \wave{6.2} observed with
the Effelsberg telescope enabled the calculation of Faraday rotation measures
($RMs$) between \wave{6.2} and the previously obtained \wave{11.1} data, which
confirmed the $ASS$ field pattern \citep{berkhuijsen03}. Combined with another
polarization survey at \wave{20.5} observed with the Very
Large Array (VLA) D-array and the Effelsberg telescope \citep{beck98},
a detailed model of the magnetic field was constructed by
\citet{fletcher04}. The spiral pitch angles were found to vary radially
from $-19\degr$ around 9\,kpc radius to $-8\degr$ around 13\,kpc radius.
The radial field component is directed \textbf{inwards}\ everywhere.
Only the mean-field $\alpha$--$\Omega$ dynamo (hereafter referred
to as mean-field dynamo) is able to generate a large-scale spiral field
that is coherent over the whole galaxy \citep{beck96}.

%The magnetic patterns of many other spiral galaxies require a superposition of several modes \citep{fletcher10}.

Measurements of the Faraday rotation of the polarized emission from 21
background sources at 1.365\,GHz and 1.652\,GHz with the VLA B-array gave
further support to the ASS field pattern and indicated that this regular
field may extend to even 25\,kpc radius \citep{han98}, but more sources are
needed for a statistically safe result \citep{stepanov08}.

The central region of M~31, which has an inclination of about $43\degr$
\citep{melchior11}, was observed by \citet{giessuebel14} at 4.86\,GHz and
8.46\,GHz with the VLA D-array and was combined with Effelsberg data at
similar frequencies. These authors detected a regular field within 0.5\,kpc
radius that is different from that in the disc.  It also reveals an $ASS$
pattern, but the magnetic pitch angle of about $-33\degr$ is much larger
than that of the disc field, and its radial field component is
directed  \textbf{outwards}, opposite to that in the disc. The central
region is known to be physically decoupled from the disc
\citep[e.g.][]{jacoby85}, with a different inclination.

Numerical models of evolving dynamos
\citep[e.g.][]{hanasz09,moss12} demonstrated how the field coherence scale
grows with galaxy age. Large-scale field reversals may still exist in
present-day galaxies if the dynamo is slow or disturbed by gravitational
interactions, while some galaxies like M~31 have reached full coherence
\citep{arshakian09}, except for the central region that probably
drives an independently operating mean-field dynamo.

%No alternative model has been proposed so far. Concepts based on primordial
%fields wound up by differential rotation \citep{sofue87b,howard97} or by
%vertical shear \citep{henriksen18,nixon18} cannot avoid field reversals
%and hence cannot explain a coherent field pattern such as observed in M~31.
%In MHD models large-scale fields are generated that are coherent over
%several kiloparsecs \citep[e.g.][]{machida13,pakmor18}, but with frequent
%azimuthal and radial reversals and hence without field coherence over the whole
%galaxy.
%%----------------------------------------

%FIG1 11cm I SS
\begin{figure*}[htbp]
\begin{center}
\includegraphics[width=12cm]{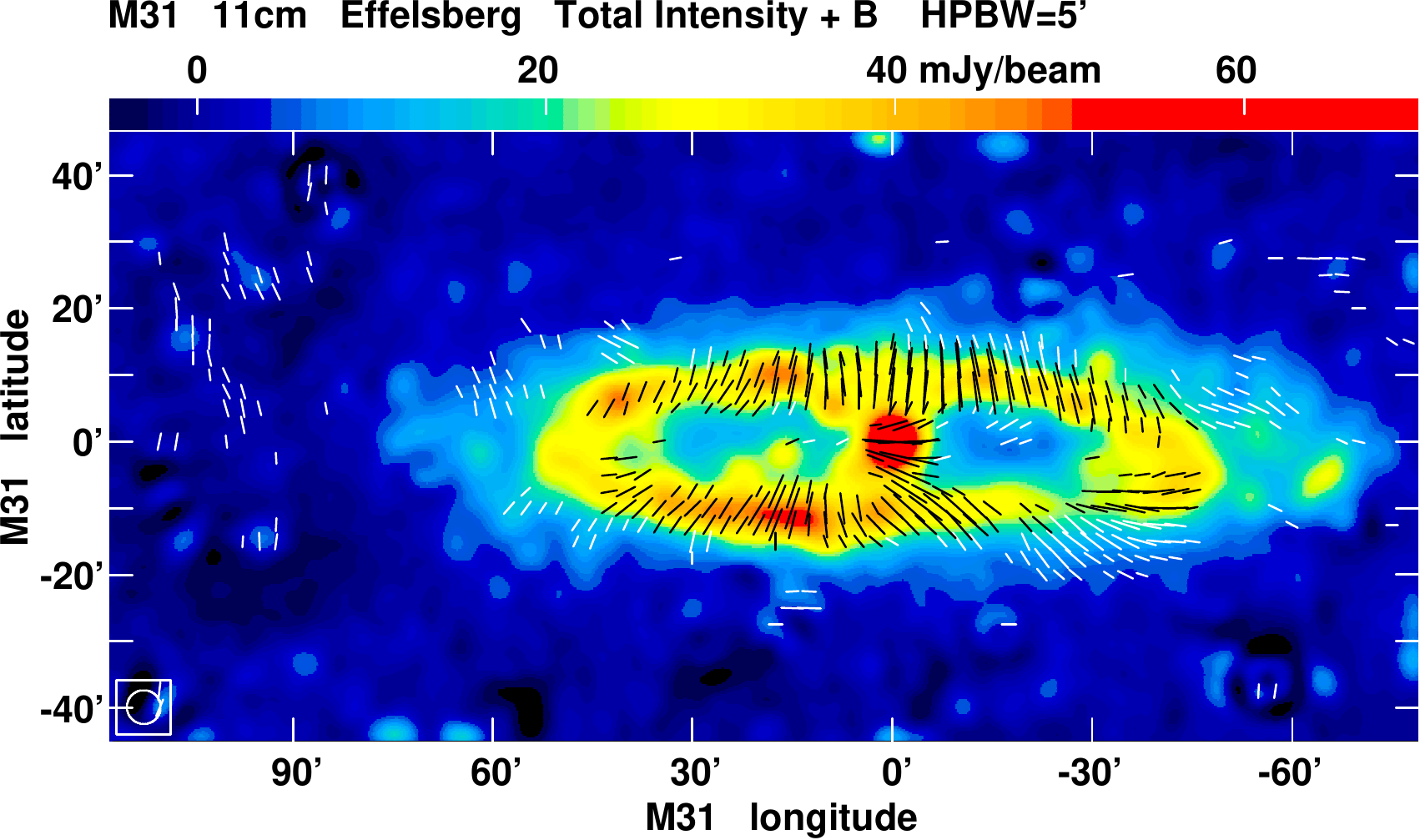}
\hfill
\caption{Total intensity $I$ (in colour) of M~31 at \wave{11.3}
smoothed to $5\arcmin$ HPBW, in coordinates along the major and minor
axes of the ring of M~31.
The rms noise is 1.0\,mJy/beam. The lines show the
apparent magnetic field orientations (not corrected for Faraday rotation)
at the same resolution with lengths proportional to polarized intensity $PI$,
where the length of a beam width corresponds to 3\,mJy/beam. No lines are plotted where
$PI$ is below 1.0\,mJy/beam or where $I$ is negative. Background sources have been
subtracted. The HPBW is indicated in the bottom left corner.}
\label{fig:cm11i}
\end{center}
\end{figure*}
%%----------------------------------------

%%----------------------------------------
%FIG2 11cm PI SS
\begin{figure*}[htbp]
\begin{center}
\includegraphics[width=12cm]{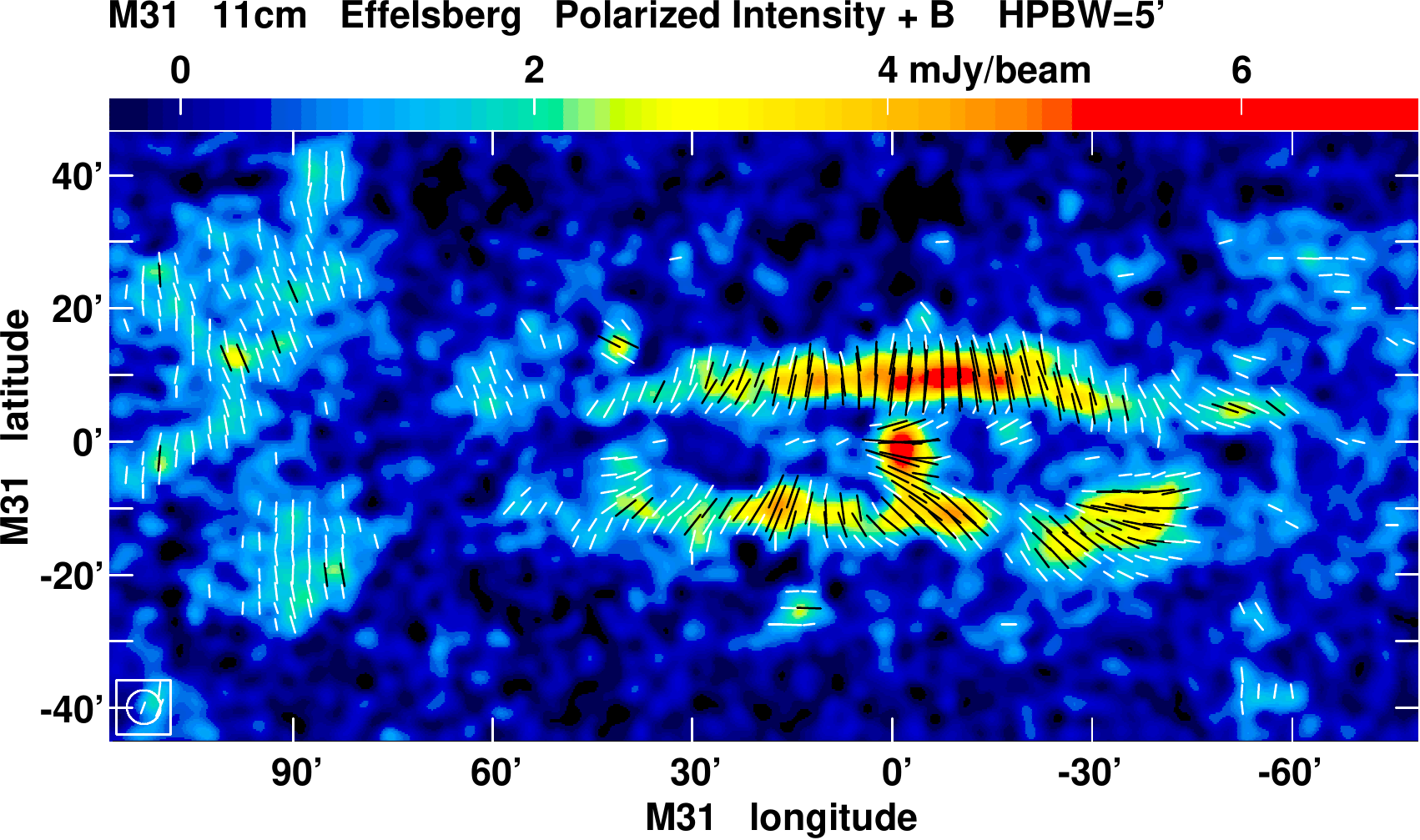}
\hfill
\caption{Polarized intensity $PI$ (in colour) of M~31 at \wave{11.3}
  smoothed to $5\arcmin$ HPBW, in coordinates along
  the major and minor axes of M~31. The rms noise is 0.3\,mJy/beam.
  The lines show the apparent magnetic field orientations (not corrected for
  Faraday rotation) at the same
  resolution with lengths proportional to polarized intensity $PI$, where
  the length of a beam width corresponds to 3\,mJy/beam. No lines are plotted where $PI$
  is below 1.0\,mJy/beam. Polarized background sources have been
  subtracted. The HPBW is indicated in the bottom left corner.}
\label{fig:cm11pi}
\end{center}
\end{figure*}
%%----------------------------------------

%%----------------------------------------
%FIG3 11cm B + Herschel
\begin{figure*}[htbp]
\begin{center}
\includegraphics[width=12cm]{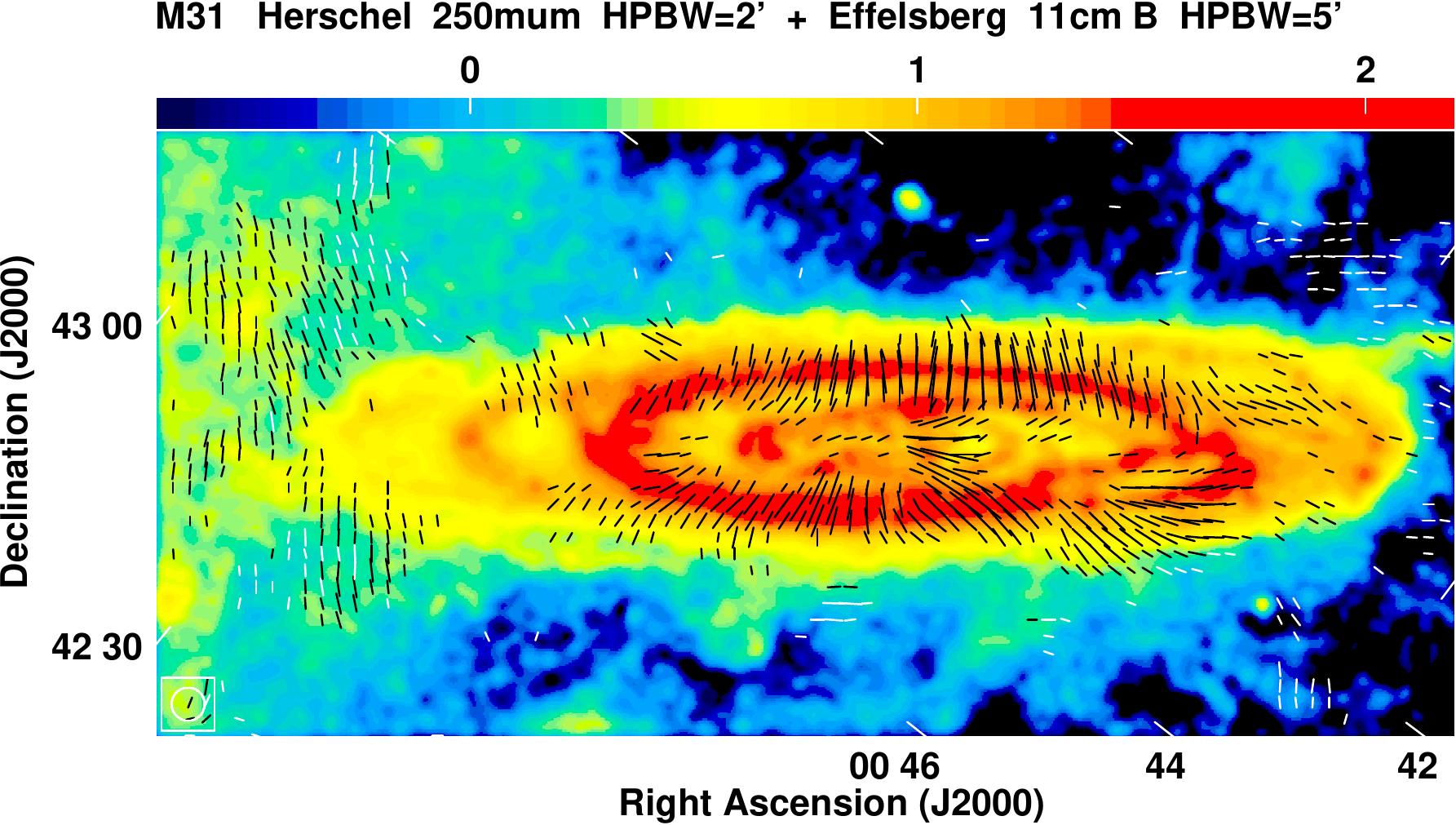}
\hfill
\caption{Apparent magnetic field orientations (not corrected for Faraday rotation)
  of M~31 at \wave{11.3} at $5\arcmin$ HPBW with lengths proportional to polarized
  intensity $PI$,
  overlaid onto an image of $250\,\mu$m infrared emission from \citet{fritz12}
  (arbitrary units, in log$_\mathrm{10}$ scale), smoothed to $2\arcmin$ HPBW,
  indicated in the bottom left corner.
  The coordinate system is rotated by $-53\degr$.
  }
\label{fig:cm11ir}
\end{center}
\end{figure*}
%%----------------------------------------

%%----------------------------------------
%FIG4 11cm PI Corr SS
\begin{figure*}[htbp]
\begin{center}
\includegraphics[width=12cm]{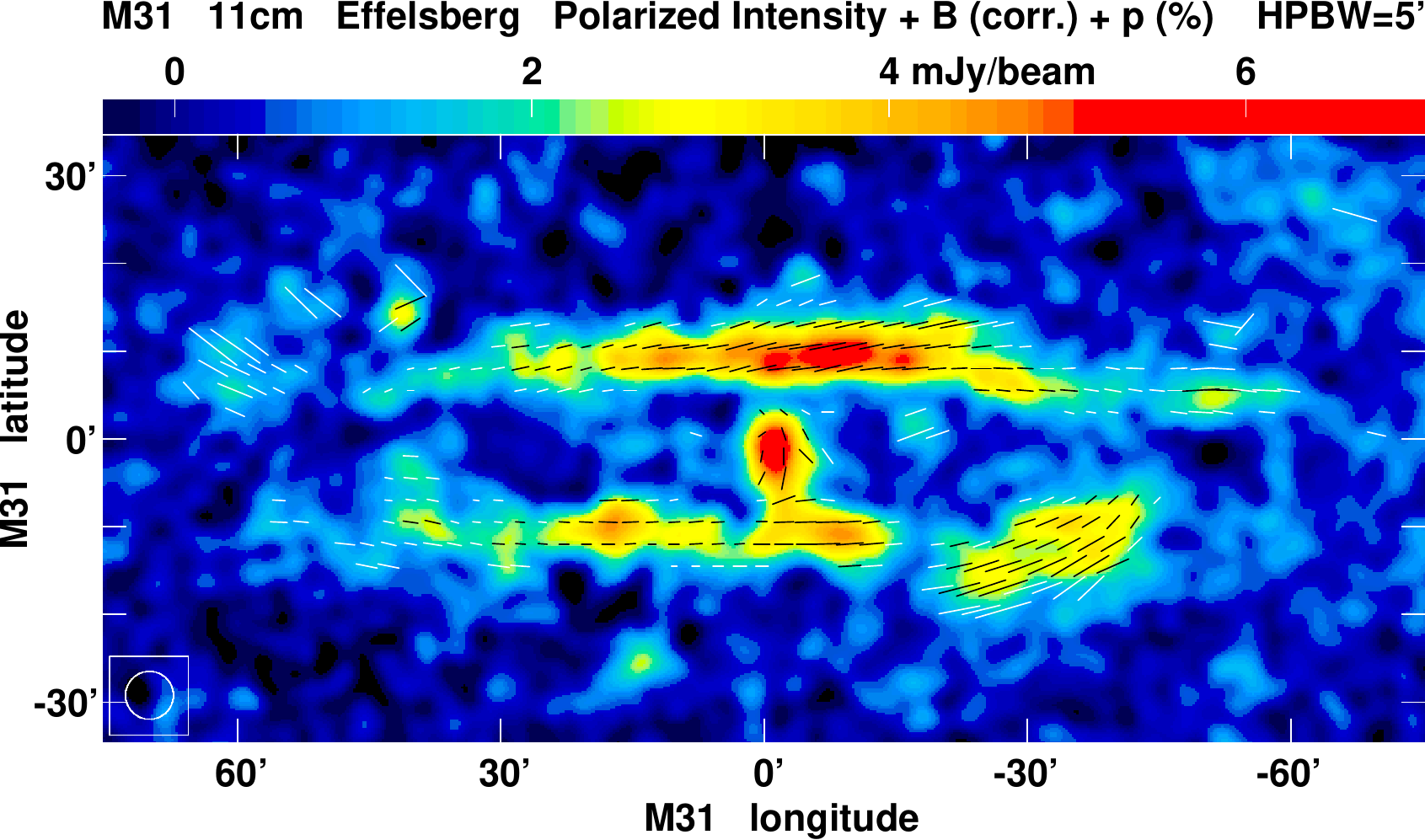}
\hfill
\caption{Polarized intensity $PI$ (in colour) of M~31 at \wave{11.3}
  smoothed to $5\arcmin$ HPBW, in coordinates along
  the major and minor axes of M~31. The rms noise is 0.3\,mJy/beam.
  The lines show the \textbf{intrinsic}\ magnetic field orientations at the same
  resolution, corrected for Faraday rotation measured between \wave{11.3}
  and \wave{6.2} (see Fig.~\ref{fig:rm6_11}),
  with lengths proportional to the degree of polarization, where
  the length of a beam width corresponds to 30\%.
  No lines are plotted where $I$ or $PI$ is below 3 times the rms noise.
  Polarized background sources have been subtracted.
  The HPBW is indicated in the bottom left corner.}
\label{fig:cm11pi2}
\end{center}
\end{figure*}
%%----------------------------------------

Polarized emission from strongly inclined galaxies like M~31 at wavelengths
above about \wave{6} is diminished by Faraday depolarization along the
line of sight through the disc by a magneto-ionic medium \citep{sokoloff98}.
Faraday depolarization increases strongly with increasing wavelength.
In order to reduce depolarization, observations at high frequencies
are desired, which also provide higher angular resolution. Furthermore,
the extension of magnetic fields into the outer disc of M~31 and a
potential radio halo should be investigated by surveys with higher
sensitivity. As synchrotron intensity decreases with decreasing wavelength,
the range \wave{3}--{6}\cm\ is optimal to observe polarized synchrotron
emission from galaxy discs and spiral arms \citep{arshakian11}.

This paper presents three new radio continuum surveys of M~31 with
improved sensitivity performed with the Effelsberg 100-m telescope at
central frequencies of 2.645, 4.85, and 8.35\,GHz (\wave{11.33},
\wave{6.18}, and \wave{3.59}). The surveys at \wave{11.3} and \wave{6.2}
cover larger fields around M~31 than the previous surveys and are also
significantly deeper (lower rms noise), especially in polarized intensity
where the rms noise is not limited by confusion from weak
background sources. Our survey at \wave{3.6} is the first one of M~31 at
such small wavelengths that covers the entire galaxy.

Section~\ref{sec:obs} summarizes the observations and data reduction.
Section~\ref{sec:results} describes the final maps,
Sect.~\ref{sec:spec} the integrated spectrum and the spectral index
distributions,
Sect.~\ref{sec:scale} the radial scale lengths of the emission components,
Sect.~\ref{sec:PI} the azimuthal variation of polarized intensity,
and Sect.~\ref{sec:rm} the Faraday rotation measures and the large-scale
field pattern.
Further analysis concerning magnetic field strengths, depolarization effects,
and propagation of cosmic rays will follow in a subsequent paper.

%---------------SECTION---------2-------------------------------------

\section{Observations and data reduction}
\label{sec:obs}

%----------------Subsection------2.1-----------------------------------

\subsection{The Effelsberg survey at \wave{11.3}}
\label{sec:obs11}

At \wave{11.3}, M~31 was observed with the single-horn
secondary-focus system of the Effelsberg 100-m telescope. The system
backend splits the total bandwidth of 80\,MHz into eight channels
from 2.60\,GHz to 2.68\,GHz. The first channel (at the lowest
frequency) was affected by strong radio frequency interference (RFI)
and could not be used. The central frequency of the remaining channels
is 2.645\,GHz (\wave{11.3}). The receiver outputs are circularly polarized
signals that are transformed into signals of Stokes $I$, $Q$, and $U$.

Maps of $196\arcmin \times 92\arcmin$  were scanned in a coordinate
system with the horizontal axis oriented along the major axis of M~31
at a position angle of $37\degr$. Thirty-four coverages in eight observation
sessions were scanned alternating along the directions parallel to
the major axis $l$ and the minor axis $b$ of the ring of M~31.
All maps were offset by
$20\arcmin$ along the major axis towards the north-east in order to include
the region of the northern spiral arm located at about $110\arcmin$ from
the centre \citep{berkhuijsen77}. One coverage took about 2\,h of
observation time.

At least one of the polarized quasar radio sources, 3C138 and 3C286, was
observed in each observation session for calibration of flux density
and polarization angle. Reference values for flux densities were taken
from the VLA Calibrator Manual \citep{perley03} and from \citet{peng00},
those for polarization angle from \citet{perley13}. The calibration
factors, averaged for these two sources, were determined for each
channel separately \citep[for details see][]{mulcahy11}. The instrumental
polarization of the Effelsberg telescope emerges from the polarized
sidelobes with 0.3--0.5\% of the peak total intensity at the frequencies
of the observations presented in this paper and is lower than
the rms noise in our maps.

%%----------------------------------------
%FIG5 6cm I
\begin{figure*}[htbp]
\begin{center}
\includegraphics[width=12cm]{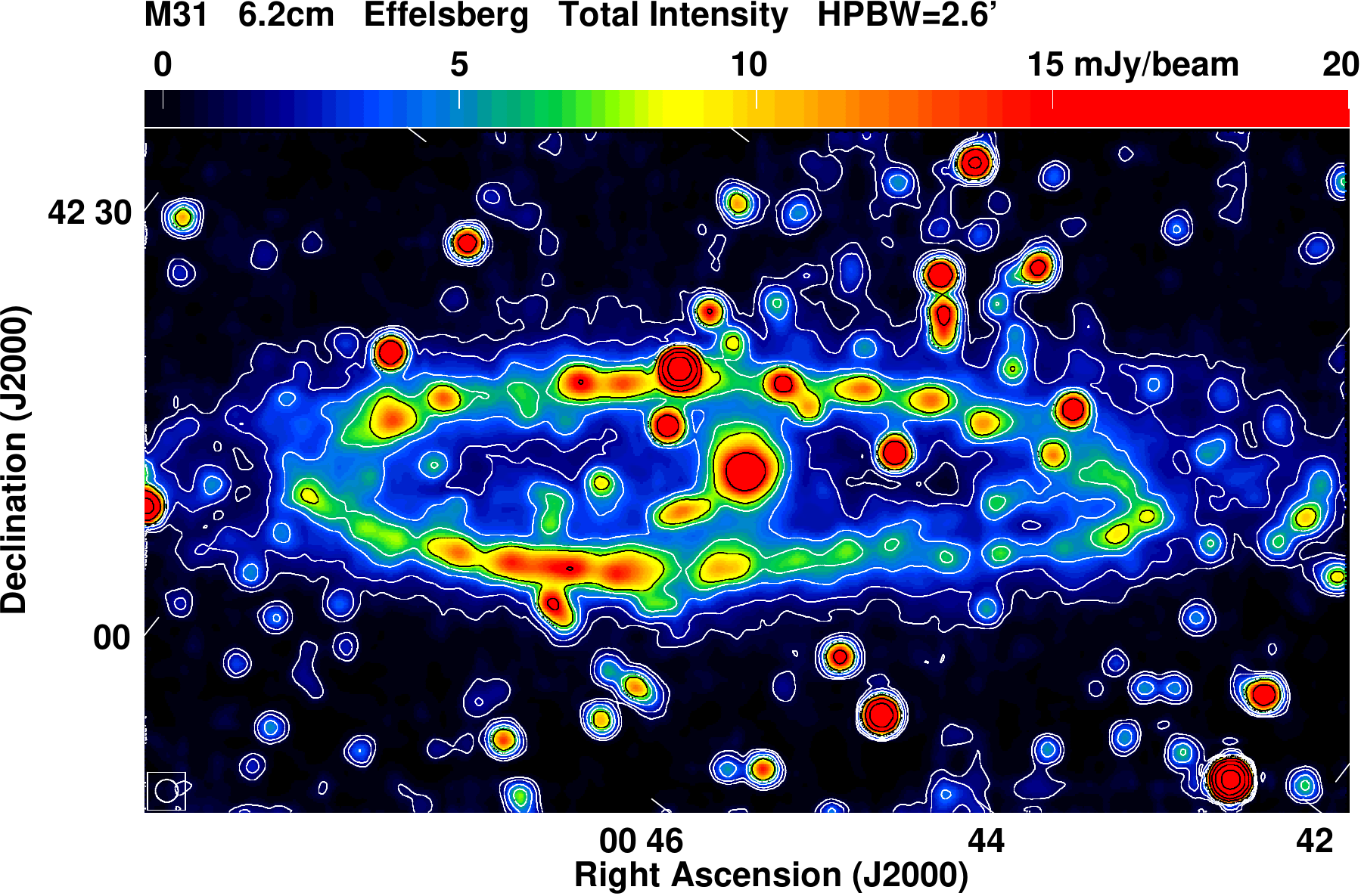}
\hfill
\caption{Total intensity $I$ (colour and contours) of M~31 at \wave{6.2}
  at the original HPBW of $2\farcm6$. Contour levels
  are at (1, 2, 4, 8, 16, 32, 64, 128) $\times$ 1\,mJy/beam. The rms
  noise is 0.3\,mJy/beam.
  The HPBW is indicated in the bottom left corner.
  The coordinate system is rotated by $-53\degr$.
  }
\label{fig:cm6i}
\end{center}
\end{figure*}
%%----------------------------------------

%%----------------------------------------
%FIG6 6cm I SS
\begin{figure*}[htbp]
\begin{center}
\includegraphics[width=12cm]{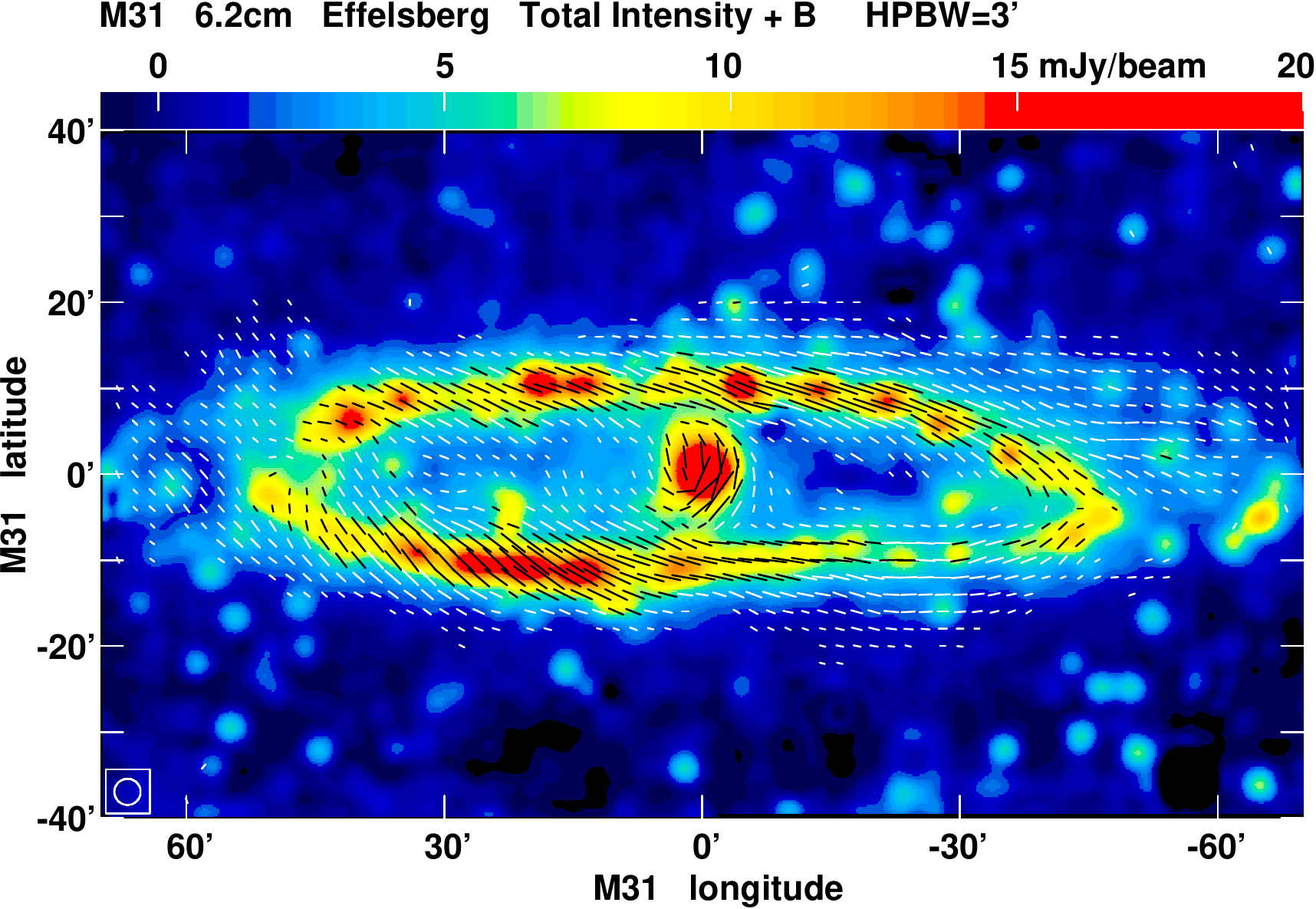}
\hfill
\caption{Total intensity $I$ (in colour) of M~31 at \wave{6.2}
  smoothed to $3\arcmin$ HPBW, in coordinates along the major and minor
  axes of M~31. The rms noise is 0.35\,mJy/beam. The lines show the
  apparent magnetic field orientations (not corrected for Faraday rotation)
  at the same resolution with lengths proportional to polarized intensity
  $PI$, where the length of a beam width corresponds to 1.5\,mJy/beam.
  No lines are plotted where $PI$ is below 0.3\,mJy/beam or where $I$ is negative.
  Background sources have been subtracted.
  The HPBW is indicated in the bottom left corner.}
\label{fig:cm6i2}
\end{center}
\end{figure*}
%%----------------------------------------

%%----------------------------------------
%FIG7 6cm PI SS
\begin{figure*}[htbp]
\begin{center}
\includegraphics[width=12cm]{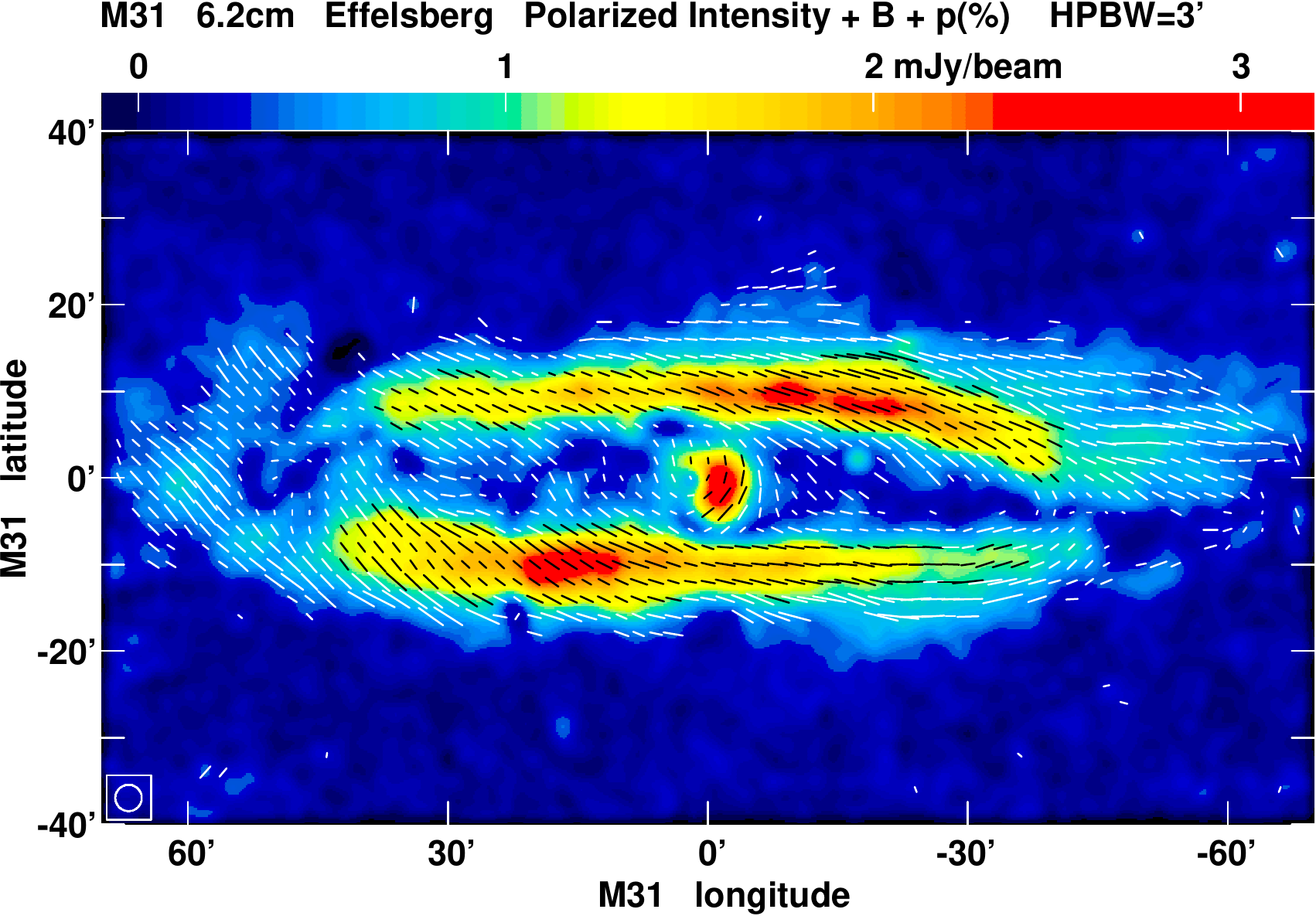}
\hfill
\caption{Polarized intensity $PI$ (in colour) of M~31 at \wave{6.2}
  smoothed to $3\arcmin$ HPBW, in coordinates along the major
  and minor axes of M~31. The rms noise is 0.06\,mJy/beam. The lines show the
  apparent magnetic field orientations (not corrected for Faraday rotation)
  at the same resolution with lengths proportional to the
  degree of polarization, where the length of a beam width corresponds to 30\%.
  No lines are plotted where $I$ or $PI$ is below 3 times the rms noise.
  Polarized background sources have been subtracted.
  The HPBW is indicated in the bottom left corner.}
\label{fig:cm6pi}
\end{center}
\end{figure*}
%%----------------------------------------

%%----------------------------------------
%FIG8 6cm NTH+B-corrected SS
\begin{figure*}[htbp]
\begin{center}
\includegraphics[width=12cm]{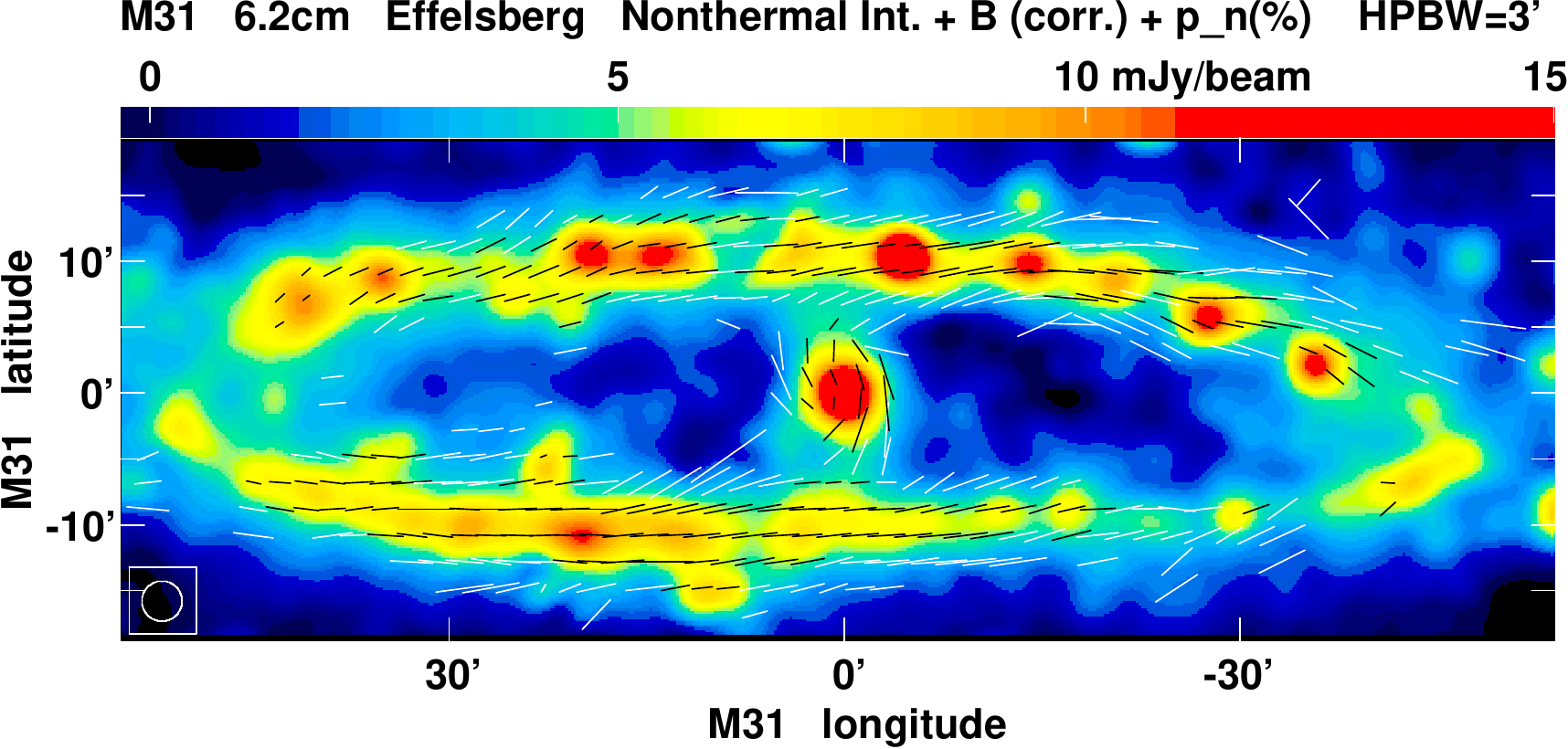}
\hfill
\caption{Total non-thermal intensity (in colour) of M~31 at \wave{6.2}
  smoothed to $3\arcmin$ HPBW, in coordinates along the
  major and minor axes of M~31. The rms noise is 0.35\,mJy/beam.
  The lines show the \textbf{intrinsic}\ magnetic field orientations
  at the same resolution, corrected for Faraday rotation
  measured between \wave{6.2} and \wave{3.6} (see Fig.~\ref{fig:rm3_6}),
  with lengths proportional to the degree of non-thermal polarization,
  where the length of a beam width corresponds to 30\%.
  No lines are plotted where $PI$ is below 0.2\,mJy/beam. Polarized
  background sources have been subtracted.
  The HPBW is indicated in the bottom left corner.}
\label{fig:cm6nth}
\end{center}
\end{figure*}
%%----------------------------------------

%%----------------------------------------
%FIG8 6cm B-corrected SS
%\begin{figure*}[htbp]
%\label{fig:cm6pi}
%\begin{center}
%\includegraphics[width=12cm]{EPS/m31cm6pi+B_180_ss_corr.eps}
%\hfill
%\caption{Polarized intensity (in colour) of M~31 at \wave{6.2} at $3\arcmin$ resolution. The rms noise is 60\,$\mu$Jy/beam.
%The lines show the intrinsic $B$ orientations, corrected for Faraday rotation (see Fig.~\ref{fig:???}), with lengths proportional to polarized intensity $PI$,
%where $10\arcmin$ correspond to 5\,mJy/beam. No lines are plotted where $PI$ is below 0.2\,mJy/beam.
%Polarized background sources have been subtracted.}
%\end{center}
%\end{figure*}
%%----------------------------------------

Data processing was performed with the {\tt NOD3}\ software package
\citep{mueller17b}. Four  of the 34 coverages could not be
used because
of strong scanning effects due to bad weather conditions. In the remaining
coverages, regions with RFI were blanked by hand. The remaining coverages,
separated into maps scanned in $l$ and in $b$, were averaged with a median
filter. The resulting maps, two each in $I$, $Q$, and $U$, were combined
in the image plane with the {\tt Mweave}\ option. This `basket-weaving'
technique reduces the scanning effects in the coverages. It was originally
developed by Chris Salter \citep[see][]{sieber79}.

The resulting maps in $Q$ and $U$ were combined into a map of polarized
intensity ($PI$) and polarization angle with the {\tt PolInt}\ option
that includes a correction for positive bias due to noise and ensures
that the $PI$ map has the same (Gaussian) noise statistics as the maps
in $Q$ and $U$ \citep{mueller17a}. Hence, we give only the noise values
for $PI$ in Table~\ref{tab:surveys}.

As we are interested in the diffuse emission from M~31, we subtracted 56
unresolved background sources above the flux density level of 10\,mJy in
total intensity ($I$) with the {\tt Gaus2}\ option of {\tt NOD2}.
In polarized intensity ($PI$), five
background sources above five times the rms noise level were detected and
subtracted. Lastly, we smoothed all the final maps from the original
resolution of $4\farcm4$ HPBW to $5\arcmin$ HPBW in
order to increase the signal-to-noise ratio.
%(Figs.~\ref{fig:cm11i} and \ref{fig:cm11pi})
The rms noise in the $4\farcm4$ image in $I$ is
about two times lower than that in the previous image at a similar frequency
\citep{beck80}, while the improvement in $PI$ is about a factor of three.

The plots of the final maps were performed with the {\tt AIPS}\ software package.

%-----------------------Subsection-------2.2---------------------------

\subsection{The Effelsberg survey at \wave{6.2}}
\label{sec:obs6}

At \wave{6.2}, M~31 was observed with the two-horn secondary-focus
system of the Effelsberg 100-m telescope. The system backend
records signal in a single band of 500\,MHz width, with a central
frequency of 4.85\,GHz (\wave{6.2}). The receiver outputs are circularly polarized
signals that are transformed into signals of Stokes $I$,
$Q$, and $U$ in a digital correlator.

Maps of $140\arcmin \times 80\arcmin$, centred on the nucleus of  M31,
were scanned in the coordinate system oriented along the major
axis. A total of 101 coverages in 20 observation sessions were scanned alternating
along the directions parallel to the major axis $l$ and the minor axis $b$. One coverage
took about 2\,h of observation time. At least one of the calibration
sources,   3C138 or 3C286, was observed in each observation session.

%%----------------------------------------
%FIG9 3cm I SS
\begin{figure*}[htbp]
\begin{center}
\includegraphics[width=12cm]{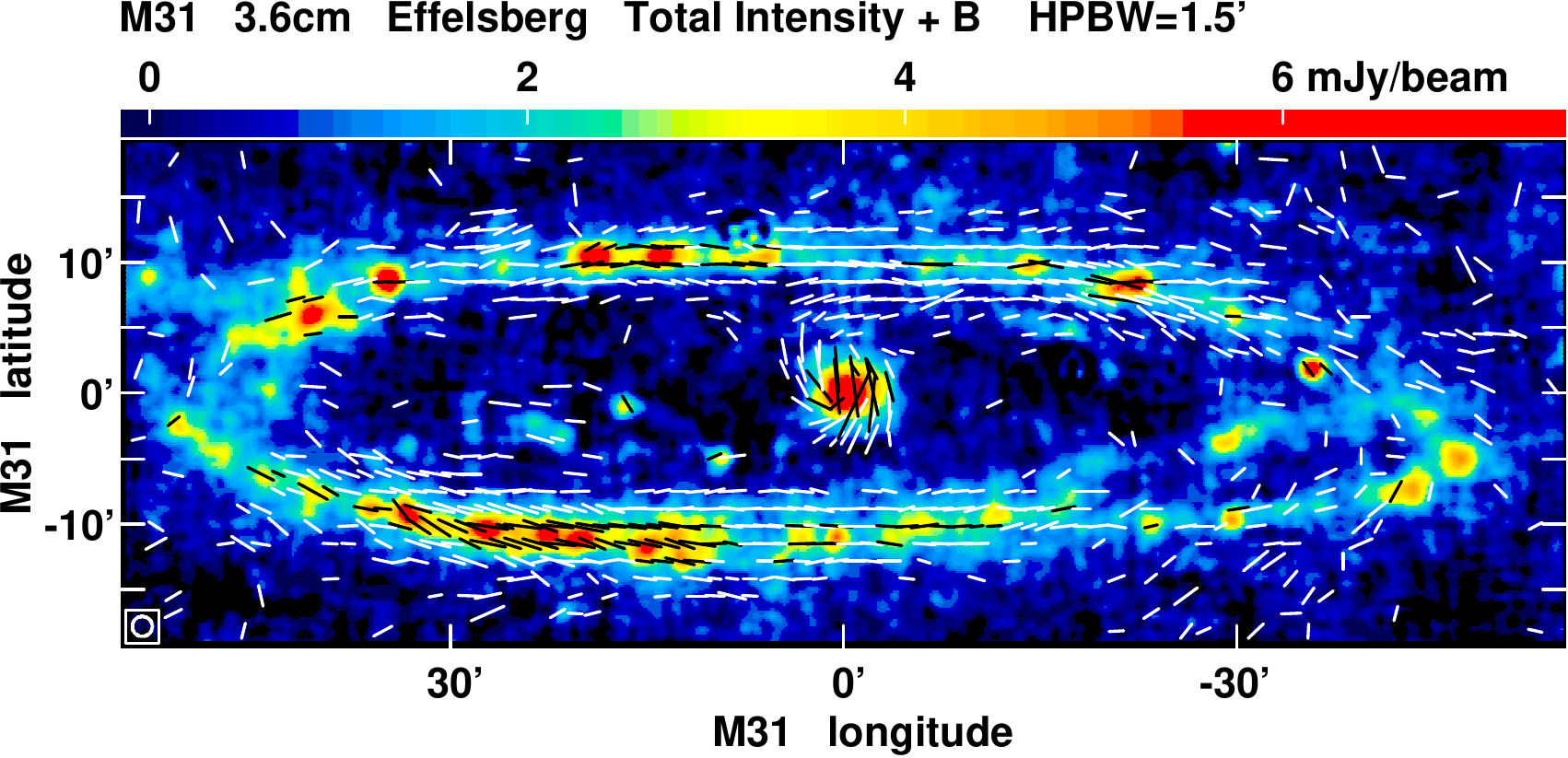}
\hfill
\caption{Total intensity $I$ (colour) of M~31 at \wave{3.6}
  smoothed to $1\farcm5$ HPBW, in coordinates along the
  major and minor axes of M~31. The rms noise is 0.25\,mJy/beam in
  the inner part ($40\arcmin \times 40\arcmin$) and 0.3\,mJy/beam in the
  outer parts. The lines show the
  apparent magnetic field orientations (not corrected for Faraday rotation)
  at the same resolution with lengths proportional to polarized intensity
  $PI$, where the length of a beam width corresponds to 0.225\,mJy/beam.
  No lines are plotted where $PI$ is below 0.15\,mJy/beam or where $I$ is negative.
  Background sources have been subtracted.
  The HPBW is indicated in the bottom left corner.}
\label{fig:cm3i}
\end{center}
\end{figure*}
%%----------------------------------------

%%----------------------------------------
%FIG10 3cm PI SS
\begin{figure*}[htbp]
\begin{center}
\includegraphics[width=12cm]{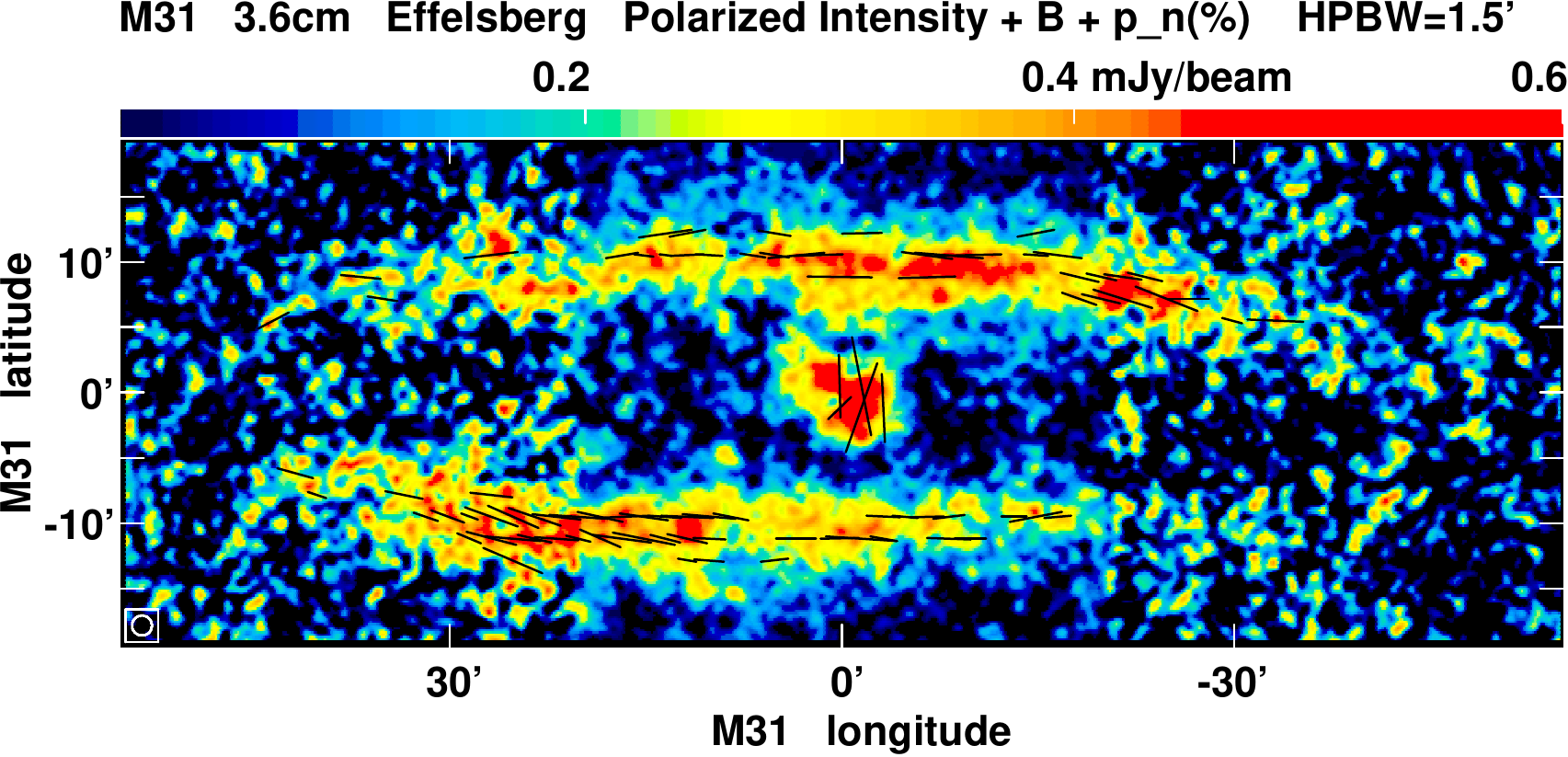}
\hfill
\caption{Polarized intensity $PI$ (colour) of M~31 at \wave{3.6}
  smoothed to $1\farcm5$ HPBW, in coordinates along the major
  and minor axes of M~31. The rms noise is 0.045\,mJy/beam in the inner part
  and 0.09\,mJy/beam in the outer parts. The lines show the
  apparent magnetic field orientations (not corrected for Faraday rotation)
  at the same resolution with lengths proportional to the degree of
  non-thermal polarization, where the length of a beam width corresponds
  to 10\%. No lines are plotted where $I$ or $PI$ is below 3 times
  the rms noise. Polarized background sources have been subtracted.
  The HPBW is indicated in the bottom left corner.}
\label{fig:cm3pi}
\end{center}
\end{figure*}
%%----------------------------------------

Thirty of the 101 coverages could not be used because of strong scanning
effects due to bad weather conditions. In the remaining coverages,
regions with RFI were blanked by hand. Data reduction was performed with
the {\tt NOD2}\ and {\tt NOD3}\ software packages. Each coverage gave
maps in Stokes $I$, $Q$, and $U$ from each of the two horns that are
separated by $8\farcm1$ in azimuthal angle on the sky. The dual-beam restoration
technique to reduce effects of weather \citep{emerson79, mueller17b} could
not be applied because scanning was done in the coordinate system of M~31
to save observation time. Hence, the coverage from the second horn
was shifted by applying the script {\tt subtrans}\
\citep[described in][]{kothes17} and added to the coverage from the
first horn. The coverages were combined applying the {\tt turboplait}\
script based on a Fourier transform \citep[following][]{emerson88} that
reduces the scanning effects by optimizing the base levels of the
orthogonally scanned coverages. All coverages were treated by applying
the {\tt NOD2}\ script {\tt fchop}\ that removes the signals at spatial
frequencies that correspond to baselines larger than the telescope
size of 100\,m. This slightly reduced the final angular resolution from
$2\farcm45$ to $2\farcm6$ HPBW.
%Details of the data reduction were described by \citet{giessuebel12}.

At \wave{6.2}, we subtracted 28 unresolved background sources above the
flux density level of 6\,mJy in $I$. In $PI$,
14 background sources above 10 times the rms noise level were
subtracted. Lastly, we smoothed the final maps to $3\arcmin$ HPBW in
order to increase the signal-to-noise ratio.
%(Figs.~\ref{fig:cm6i2} and \ref{fig:cm6pi})
The rms noise in the $2\farcm6$ image in $I$ is about
1.5 times lower than that in the previous image at the same
frequency \citep{berkhuijsen03}, while the improvement in $PI$ is a factor
of about 2.5.

%------------------------Subsection----2.3-----------------------

\subsection{The Effelsberg survey at \wave{3.6}}
\label{sec:obs3}

At \wave{3.6}, M~31 was observed with the single-horn secondary-focus
system of the Effelsberg 100-m telescope with a total size of
$116\arcmin \times 40\arcmin$. The system backend
records signal in a single band of 1100\,MHz width, with a
central frequency of 8.35\,GHz. The receiver outputs are circularly
polarized signals that are transformed into signals of Stokes $I$,
$Q$, and $U$ in a digital correlator.

The project required a huge effort due to the weak emission of the galaxy
at this frequency and the small telescope beam compared to the large
angular size of M~31. We started with a small test field of
$10\arcmin \times 10\arcmin$ centred on the galaxy nucleus, taking
about 25\,min observation time per coverage, followed by two adjacent
fields of $25\arcmin \times 17\arcmin$ to the north-west and south-east
and another two fields of $17\arcmin \times 25\arcmin$ to the north-east
and south-west, taking about 1\,h observation time per coverage and per
field. Overlaps of $2\arcmin$ allowed us to adjust the base levels.
Between December 2001 and October 2007, about 40 coverages per field were scanned
along alternating directions parallel and perpendicular to the major axis
(galaxy coordinates $l$ and $b$). At least one of the two calibration sources,
3C138 or 3C286, was observed in each observation session.

Data reduction was performed with the {\tt NOD2}\ and {\tt NOD3}\ software packages.
About 10\% of the coverages could not be used because of bad weather
conditions. In the remaining coverages, regions with RFI were blanked by
hand. The coverages from each field scanned in the  $l$ and $b$ directions
were averaged separately. Combining the $l$ and $b$ maps from each
field of the inner region needed special attention because the diffuse
emission is more extended in the $l$ direction than the field size, so that
the option {\tt turboplait}\ or {\tt Mweave}\ could not be applied.
The following procedure was performed:
(1) In the $b$ map (where strong unresolved sources are subtracted), fit
and subtract linear baselines in the $l$ direction;
(2) Subtract the resulting $b$ map from the original $b$ map to obtain a
map of large-scale structures in the $b$ direction;
(3) In the $l$ maps (where again strong unresolved sources are subtracted),
fit and subtract linear baselines in the $l$ direction;
(4) Subtract the resulting $l$ map from the original l map to obtain a map
of large-scale structures in the $l$ direction;
(5) Compute the difference of the maps obtained in steps (2) and (4) to
obtain the large-scale emission missing in the $l$ map;
(6) Add the result from step (5) to the $l$ map.
Details of the data reduction  have been described by \citet{giessuebel12}.

The combination of the first five fields yielded a map of $40\arcmin \times
40\arcmin$ centred on the galaxy nucleus.
The high investment in observation time resulted in a low rms noise of
0.25\,mJy/beam in $I$ and 0.06\,mJy/beam in $PI$.

After successful completion of the central part, we decided to extend the
survey to cover most of the galaxy disc out to a distance of $l=\pm58\arcmin$ (13.2\,kpc)
 from the centre. This was achieved by observing two fields
on both sides of the major axis
of $40\arcmin \times 40\arcmin$ each, overlapping by $2\arcmin$ with the
central part, taking about 1.7\,h of observation time per coverage.
Between November 2011 and September 2012, 24 coverages per field were scanned
alternating along the horizontal ($l$) and vertical ($b$) galaxy
coordinates, of which ten could not be used because of bad weather
conditions. The combination of the fields was done by adjusting the base levels
in the overlap regions. The seven fields cover a total size of
$116\arcmin \times 40\arcmin$ in the coordinate system of M~31.
As the number of coverages of the two outer fields was smaller than  for
the inner part, the rms noise is higher, i.e. 0.3\,mJy/beam in $I$ and
0.12\,mJy/beam in $PI$.

The extent of the final map of $b=\pm20\arcmin$ from the major axis
of M~31 (17.5\,kpc in the galaxy plane)
was chosen to keep the total observation time within a manageable
limit, even though we were aware that some of the faint diffuse emission from the outer
disc would be missing due to the base level subtraction.

In an attempt to correct for the largest scales of emission, we observed a
grid of 11 scans of $70\arcmin$ in length perpendicular to the plane (in $b$),
separated by $4\arcmin$ along the plane, and combined them into an
undersampled map of $40\arcmin \times 70\arcmin$.
%However, it turned out that these scans were too long to ensure subtraction of the base level by linear interpolation to a precision required to detect the extended
%low-level emission. The reason could have been ground radiation that slowly varies with elevation of the telescope and hence with time.
%However, only half of the observed scans could be used. The other half was heavily affected by variations of the atmospheric emission with elevation:
All observations took place before local sidereal time
$LST=00^\mathrm{h} 40^\mathrm{m}$ (i.e. before M~31 crossed the meridian).
When the galaxy was rising, its major axis remained
approximately parallel to the Earth's horizon. Our scanning speed of
$20\arcsec$/s was close to the apparent speed of the sky. When scanning in the
negative $b$ direction before $LST=00^\mathrm{h} 40^\mathrm{m}$, the
telescope remained almost still as the sky moved across. The atmospheric
and ground emission thus remained more or less constant during a scan.
On the other hand, when scanning in the positive $b$ direction, the telescope
had to move fast in elevation, almost two times as fast as just tracking M~31,
so that the signal was affected by varying atmospheric and ground emission.
In order to add as little uncertainty to the correction as possible, we
decided to use only those scanned in the negative $b$ direction where the
effects of the atmospheric and ground emission were negligible.

To ensure well-defined base levels for the correction grid, we observed
three additional $l$ scans of $70\arcmin$ length crossing the
lower ends of the $b$ scans and three scans crossing their upper
ends. A comparison with the levels of the $b$ scans with those of the $l$
scans at the intersection points showed that only minor corrections were
necessary. This demonstrated that the $b$ scans of the grid were long
enough to define correct base levels.

The correction grid was then used to define the base level of each of the
five inner fields. The amount of missing flux at the map edges compared
to the correction grid was determined by subtracting the average over five
pixels in the scanning direction. The $4\arcmin$ gap between the correction
scans was then interpolated by fitting a polynomial to these data points.
The order of the polynomial was chosen to be  as small as possible, from first
to fourth, depending on which one best described the shape determined by
the grid. The actual baseline correction along the direction of the grid
used only a linear fit between the bottom and top to ensure an unbiased
correction. Only small and linear additional baseline corrections were
applied where necessary so that the individual maps matched up correctly
in the overlap region. A more thorough description and comparison of the
maps before and after
the correction can be found in \citet{giessuebel12}.

At \wave{3.6}, we subtracted 38 unresolved background sources above the
flux density level of 1.2\,mJy in $I$. In $PI$,
six background sources above five times the rms noise level were detected
and subtracted. Lastly, we smoothed the final maps from the original
resolution of $1\farcm4$ to $1\farcm5$ HPBW
%(Figs.~\ref{fig:cm3i} and \ref{fig:cm3pi})
in order to increase the signal-to-noise ratios.

%-----------------------SECTION-------3------------------------

\section{Final maps}
\label{sec:results}

The total radio continuum emission from M~31 (Figures~\ref{fig:cm11i},
\ref{fig:cm6i2}, and \ref{fig:cm3i}) is concentrated in the well-known
ring-like structure between 7\,kpc and 13\,kpc from the galaxy centre.
Strong emission emerges from regions with high star formation rates ($SFRs$)
evident from their H$\alpha$ emission \citep{devereux94}, as discussed by
\citet{taba10}. The spatially resolved radio--far-infrared correlation
in galaxies, first found in M~31 \citep{beck88}, was studied in M~31 in
detail \citep{hoernes98,berkhuijsen13} and confirmed the close
relationship between $SFRs$ and total radio continuum emission.
The central region of M~31 is radio-bright in spite of its low SFR,
possibly because cosmic-ray electrons are
(re-) accelerated by shock fronts responsible for
the filamentary H$\alpha$ emission \citep{jacoby85}.

%---------------------------------------------------------------------------
%TABLE3
\begin{table*}
\begin{center}
  \caption{Integrated flux densities of M~31 for three radial ranges for
    total intensity $I$, non-thermal intensity $NTH$, and thermal
    intensity $TH$.}
  \label{tab:flux}
  \begin{tabular}{lccllcll}
  \hline
  Freq.     & $S_\mathrm{min}$    & $R<1$\,kpc        & $R<5$\,kpc        & $R<16$\,kpc             & $R<16$\,kpc       & $R<16$\,kpc            & Reference \\
  $[$MHz$]$ & [mJy] $^1$   & $I$~[mJy]      & $I$~[mJy]      & $I$~[mJy]            & $TH$~[mJy] $^2$ & $NTH$~[mJy] \\
  \hline
  327       & 47                 & $244\pm18$        &   $1320\pm100$     &   $14070\pm1030$ $^3$   &   $775\pm80$      &   $13290\pm1040$       & \citet{golla89} \\
  327       & 47                 & $183\pm18$        &   $1230\pm120$    &   $12470\pm1250$        &   $775\pm80$      &   $11690\pm1250$       & \citet{giessuebel12} \\
  408       & 40                 & $167\pm12$        &   $1190\pm90$     &   $10770\pm750$         &   $760\pm75$      &   $10010\pm760$        & \citet{beck+graeve82} \\
  1465      & 15                 & $110\pm10$        &    $~~460\pm50$   &    $~~4790\pm480$       &   $670\pm70$      &   $~~4120\pm490$       & \citet{beck98} \\
  2645      & 10                 & $47\pm5$          &    $~~354\pm35$   &    $~~3190\pm320$       &   $625\pm65$      &   $~~2570\pm330$       & This paper\\
  2702      & 10                 & $43\pm2$          &    $~~321\pm22$   &    $~~2670\pm340$       &   $625\pm65$      &   $~~2040\pm350$       & \citet{beck80} \\
  4850      & 6                  & $39\pm4$          &    $~~194\pm22$   &    $~~1850\pm210$ $^4$  &   $590\pm60$      &   $~~1260\pm220$       & \citet{berkhuijsen03} \\
  4850      & 6                  & $37\pm4$          &    $~~224\pm22$   &    $~~2010\pm200$       &   $590\pm60$      &   $~~1420\pm210$       & This paper \\
  8350      & 4                  & $28\pm3$          &    $~~~~80\pm50$  &    $~~1410\pm200$ $^5$  &   $560\pm60$      &   $~~~~850\pm210$      & This paper \\
  \hline
  \end{tabular}
  \footnotesize
\item Notes:
$^1$ Minimum flux density of subtracted sources (see text for details);
$^2$  thermal flux densities are based on the thermal map at \wave{20.5}
  of \citet{taba13b}, increased by 5\% for missing areas near the major
  axis at $R>13$\,kpc;
$^3$ corrected for missing spacings;
$^4$ corrected for missing areas near the major axis at $R>13$\,kpc;
$^5$ increased by 5\% for missing areas near the major axis at $R>13$\,kpc.
\end{center}
\end{table*}
\normalsize
%---------------------------------------------------------------------------

Figure~\ref{fig:cm6i} shows the full field of the total emission at
\wave{6.2} before subtraction of the background sources that are unrelated
to M~31. These sources have been discussed before by \citet{berkhuijsen83}.

The linearly polarized emission from M~31 (Figures~\ref{fig:cm11pi},
\ref{fig:cm6pi}, and \ref{fig:cm3pi}) is also concentrated in the ring-like
structure. The main differences to the distribution of total emission are
the minima around the major axis of the projected ring.
This shows that the ordered magnetic
field in the ring is oriented almost along the line of sight on the major
axis, and hence almost follows the ring \citep{beck82}. The variation of
polarized intensity along the ring is discussed in Section~\ref{sec:rm}.

Significant polarized emission (more than 3 times the rms noise) is detected
at \wave{11.3} also outside the ring of M~31 in the NE
(Fig.~\ref{fig:cm11pi}).
Some regions are about the same size as the telescope beam and could be weak
polarized background sources. Extended polarized patches  may originate
in the foreground of our Milky Way. Polarized patches in the region of M~31
were also observed  at \wave{21.1}
\citep{berkhuijsen03}. They are called Faraday ghosts, and are caused by Faraday rotation and
depolarization of smooth polarized emission occurring in
nearby magnetized regions in the foreground.
They appear only in $PI$, but not in total intensity $I$, and are
especially prominent when observing at lower frequencies, for example with the
WSRT at around \wave{90} \citep[e.g.][]{haverkorn03,schnitzeler09}
or around $\lambda2\,\mathrm{m}$ (150\,MHz) \citep[e.g.][]{iacobelli13}, or with
the Low Frequency Array (LOFAR)
at around $\lambda2\,\mathrm{m}$ \citep[e.g.][]{jelic15,vaneck17}. Some of the polarized regions outside of the ring to the north-east
may originate in high-velocity $\HI$ clouds belonging to M~31 that are mixed
with dust seen at $250\,\mu$m \citep{fritz12} (see Fig.~\ref{fig:cm11ir}).

%The large regions of $20\arcmin-60\arcmin$ size north of M~31 may
% originate in dust clouds associated with M~31 (Fig.~\ref{fig:cm11ir}). .....%...

The apparent magnetic field orientations (i.e. the polarization angles
$+90\degr$) in Figs.~\ref{fig:cm11i}, \ref{fig:cm6i2}, and \ref{fig:cm3i}
strongly differ for the three frequencies, demonstrating the action of
Faraday rotation. The orientations of the intrinsic magnetic field (corrected
for Faraday rotation) shown in Figures~\ref{fig:cm11pi2} and \ref{fig:cm6nth}
agree well and show that the magnetic field closely follows the ring. In
Sect.~\ref{sec:rm}, maps of Faraday rotation measures are discussed.

The two regions of polarized emission outside the ring towards the north,
detected at \wave{11.3} and \wave{6.2}, allowed us to compute the intrinsic
field orientations in these areas (Fig.~\ref{fig:cm11pi2}, near the left
edge of the plot). These strongly deviate from the orientation of the ring
and suggest a location in high-velocity clouds around M~31 or in the
Milky Way foreground.

%----------------------SECTION-----4--------------------------------------

\section{Spectral index}
\label{sec:spec}

\subsection{Spectral index of the integrated flux density}
\label{subsect:specint}

%%----------------------------------------
%FIG11 Integrated spectral index
\begin{figure}[htbp]
\begin{center}
\includegraphics[width=0.8\columnwidth,angle=270]{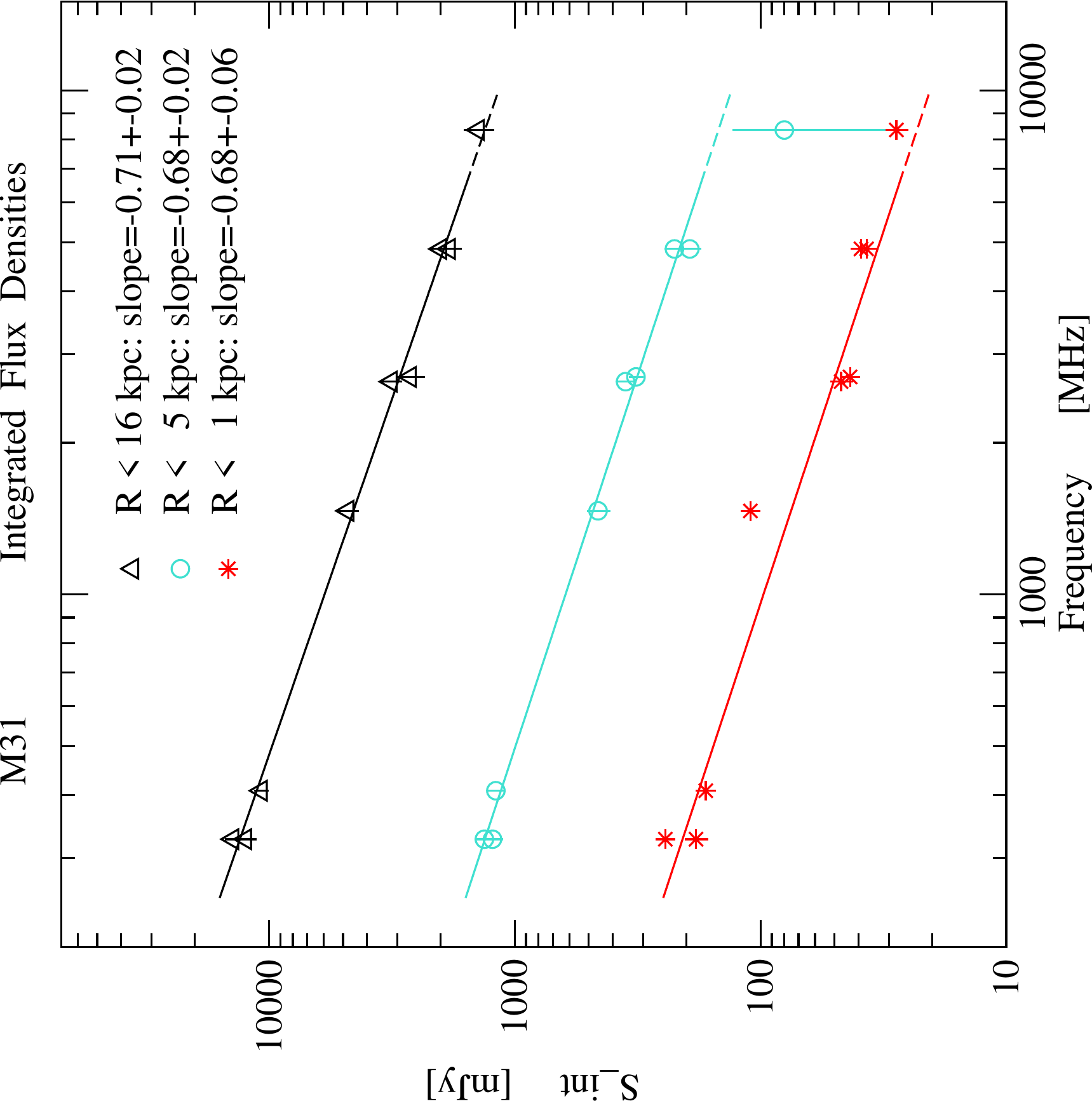}
\hfill
\caption{Spectrum of the integrated flux density of total intensity $I$
  integrated to radii of
  1\,kpc (red), 5\,kpc (cyan), and 16\,kpc (black). The slopes of the
  fitted lines are also given in the figure. The points at \wave{3.6} were
  not used for the fits.}
\label{fig:spectrum1}
\end{center}
\end{figure}
%%----------------------------------------

%%----------------------------------------
%FIG12 Integrated spectral index I, NTH, TH
\begin{figure}[htbp]
\begin{center}
\includegraphics[width=0.8\columnwidth,angle=270]{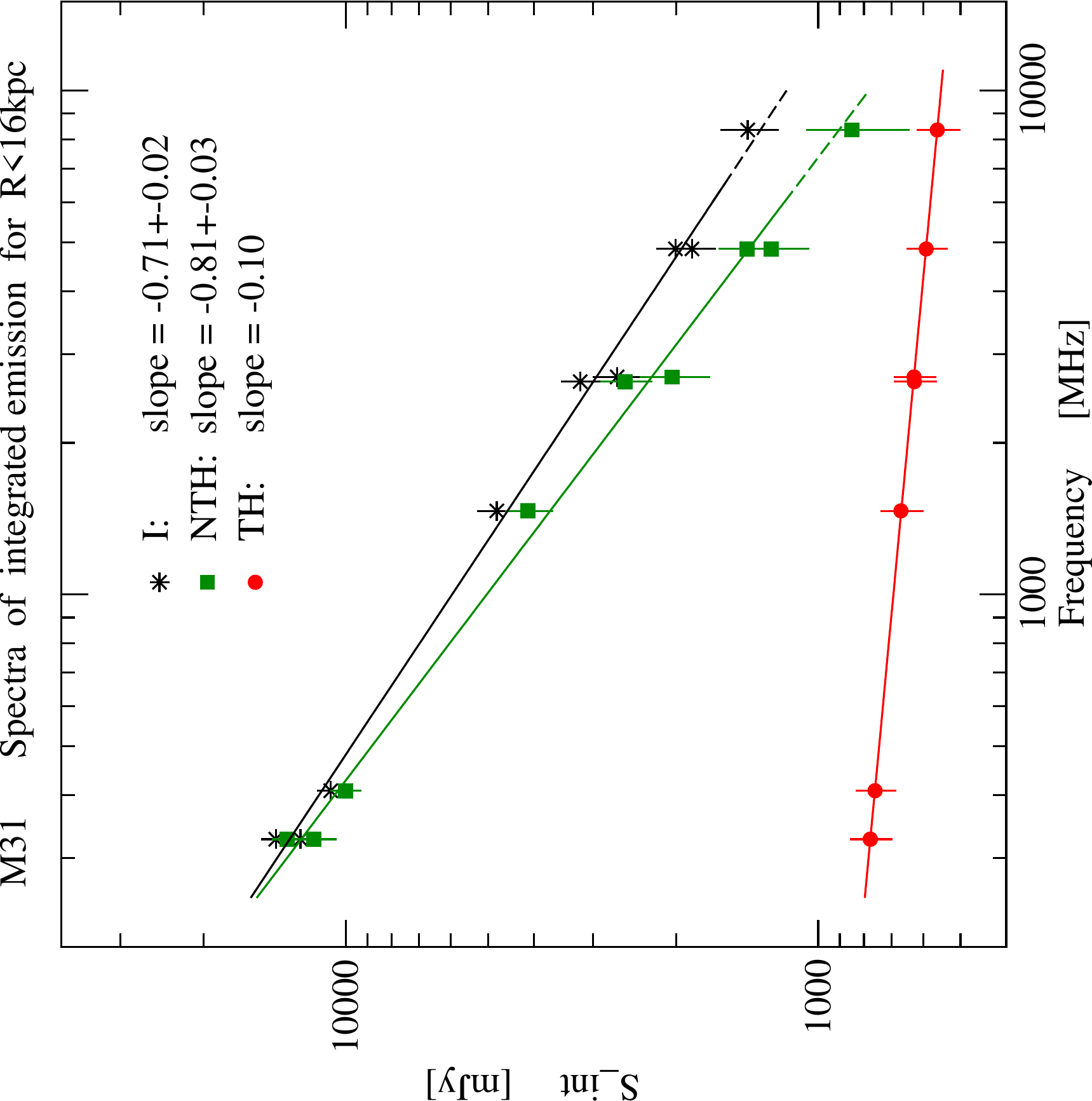}
\hfill
\caption{Spectrum of the integrated flux densities of total intensity $I$ (black),
  non-thermal intensity $NTH$ (green), and thermal intensity $TH$
  (red), integrated to a radius of 16\,kpc. The slopes of the fitted lines
  are also given in the figure. The uncertain points for $I$ and $NTH$
  at \wave{3.6} were not used for the fits. The points for $TH$
  were scaled from the thermal map at \wave{20.5} using a slope of -0.1.}
\label{fig:spectrum2}
\end{center}
\end{figure}
%%----------------------------------------

The galaxy M~31 has been mapped in radio continuum at more than six
frequencies (see Table~\ref{tab:flux}).
In order to subtract the same point sources from
all maps, the flux densities of the sources subtracted at 408\,MHz
were scaled to the frequencies of the other maps, assuming a constant
spectral index of $\alpha=0.7$ (where $S \propto \nu^{-\alpha}$).
These flux densities are indicated by $S_\mathrm{min}$ in Table~\ref{tab:flux}.
After subtracting sources with flux densities $S > S_\mathrm{min}$,
we used the data to calculate the total spectral index $\alpha$
of the emission integrated in three radial intervals
in M~31 (i.e. $R < 1$\,kpc, $R < 5$\,kpc and $R < 16$\,kpc).

Not all maps extend to $R = 16$\,kpc (= 70\arcmin) along the major axis.
By comparing them to larger maps we estimated that in these cases
about 5\% of the flux density was missing and we corrected for this.

The base levels of the maps are usually set to zero in strips that are a few beam
widths wide and parallel to the major axis at $|b| \approx 30\arcmin$. Only
the new map at \wave{3.6} does not reach that far. Therefore, we
adjusted this map to the background level of the new map at \wave{6.2}
at $|b| \approx 20\arcmin$ (after smoothing it to HPBW = $3\arcmin$) using the
task {\tt bascor}\ of the {\tt NOD2}\ system. For this adjustment we assumed
$\alpha = 0.9$ between \wave{6.2} and \wave{3.6} at $|b| \approx 20\arcmin$. This
resulted in an increase in the integrated flux density for $R < 16$\,kpc
of 200\,mJy.

The integration was done by adding the emission in circular rings
around the centre in the plane of the galaxy. The resulting flux
densities in total power of the three radial intervals are given in
Table~\ref{tab:flux} and  are shown in Figure~\ref{fig:spectrum1}.
The low value at \wave{3.6} for $R < 5$\,kpc indicates that, in spite of the
great efforts (Section~\ref{sec:obs3}), the subtracted base level of the
small map around the central region was still too high.

Because of the uncertainty in the flux densities at \wave{3.6}, we calculated
the spectral index between \wave{92} (0.327\,GHz) and \wave{6.2} (4.85\,GHz)
yielding $\alpha = 0.68 \pm 0.06$ for $R < 1$\,kpc, $\alpha = 0.68 \pm 0.02$
for $R < 5$\,kpc, and $\alpha = 0.71 \pm 0.02$ for $R < 16$\,kpc.
\citet{berkhuijsen03} found $\alpha = 0.83 \pm 0.13$ for $R < 16$\,kpc
using only data at \wave{20.5} and \wave{6.2}. Within the errors their value is
consistent with our value, but since our value is based on more data points
and a larger frequency interval, our value of $\alpha = 0.71 \pm 0.02$
supersedes the old one.

In order to derive the non-thermal spectral index of the emission
integrated to $R < 16$\,kpc, we subtracted the integrated thermal emission
from the total emission at each frequency, scaled from the
integrated thermal emission at \wave{20.5} given by \citet{taba13a},
using the spectral index of optically thin free-free emission of 0.1.
Before scaling, we increased the value at \wave{20.5} by 5\% to account
for missing areas in the thermal map near the major axis.
Figure~\ref{fig:spectrum2} shows the total, non-thermal, and thermal spectra
of the integrated emission. A weighted fit through the non-thermal flux
densities between \wave{92} and \wave{6.2} gives the non-thermal spectral
index $\alphan = 0.81 \pm 0.03$. This shows that the value of $\alphan = 1.0$
between \wave{20.5} and \wave{6.2} assumed by \citet{berkhuijsen03} was
too large, and demonstrates that spectral indices measured between
only two frequencies do not always agree with that derived by fitting the
data at many frequencies.

%------------------------------Subsection   4.2-------------------------

\subsection{Maps of spectral index between \wave{20.5} and \wave{3.6}}
\label{subsect:specmaps}

%%----------------------------------------
%FIG13 Spectral index distribution
\begin{figure*}[htbp]
\begin{center}
\includegraphics[width=12cm]{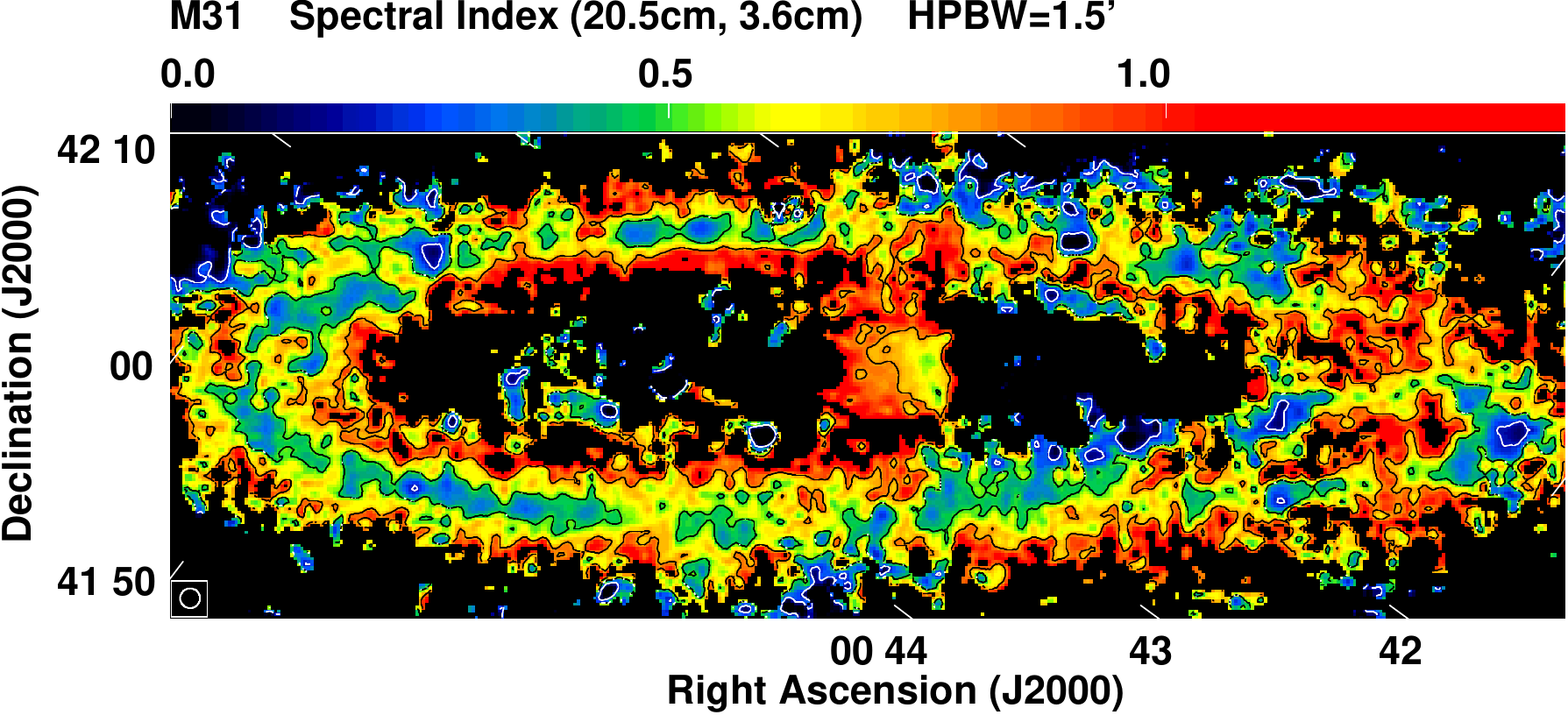}
\hfill
\caption{Spectral index of the total intensity between \wave{20.5} and
  \wave{3.6} at $1\farcm5$ HPBW,
  calculated at pixels where $I$ at both frequencies exceeds two times the rms noise.
  Contour levels are at 0.2, 0.5, and 0.8. Background sources have been subtracted.
  The HPBW is indicated in the bottom left corner.
  The coordinate system is rotated by $-53\degr$.
  }
\label{fig:alpha}
\end{center}
\end{figure*}
%%----------------------------------------

%%----------------------------------------
%FIG14 Non-thermal spectral index distribution
\begin{figure*}[htbp]
\begin{center}
\includegraphics[width=12cm]{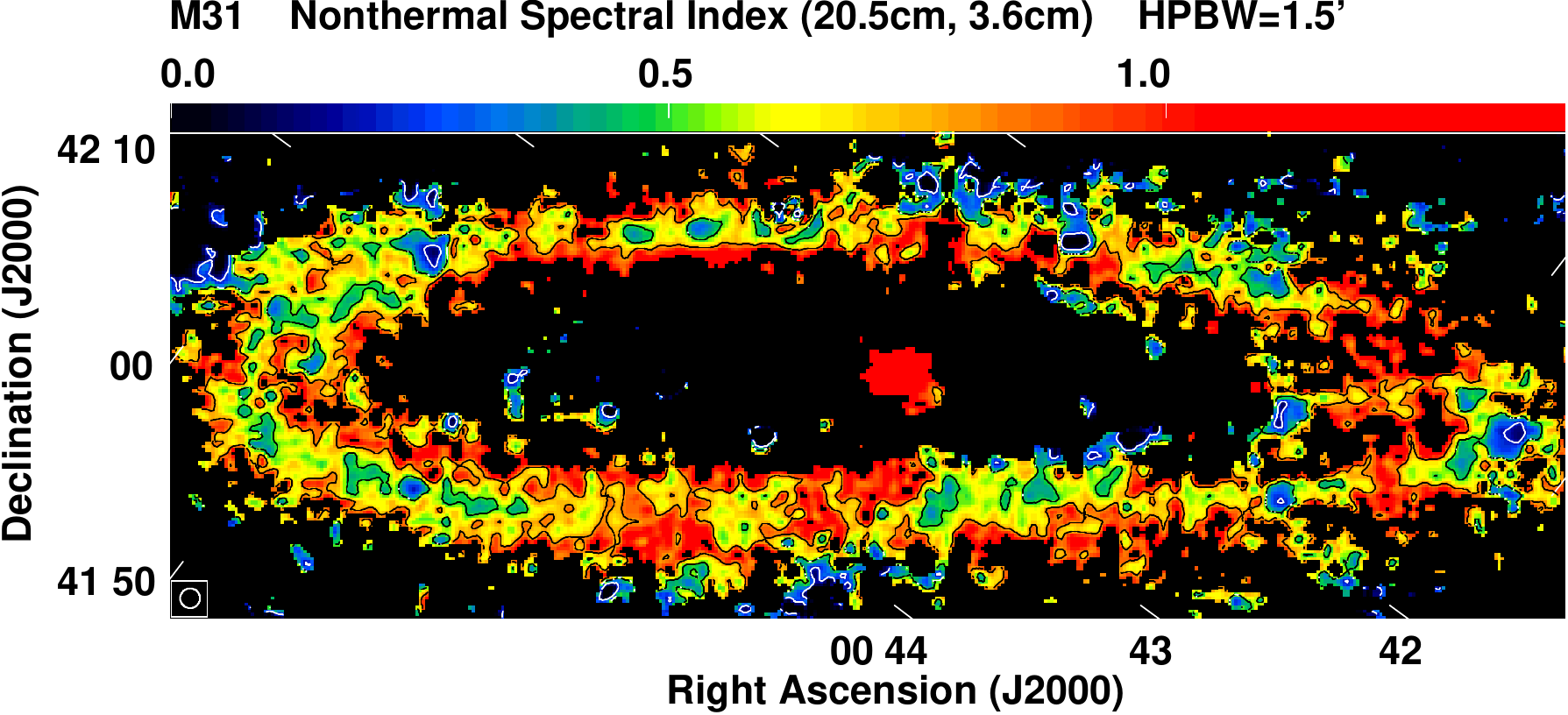}
\hfill
\caption{Spectral index of the non-thermal intensity between \wave{20.5} and
  \wave{3.6} at $1\farcm5$ HPBW,
  calculated at pixels where $I$ at both frequencies exceeds two times the rms noise.
  Contour levels are at 0.2, 0.5, and 0.8. Background sources have been subtracted.
  The HPBW is indicated in the bottom left corner.
  The coordinate system is rotated by $-53\degr$.
  }
\label{fig:alpha_nth}
\end{center}
\end{figure*}
%%----------------------------------------

Figure~\ref{fig:spectrum2} shows that the integrated flux densities for
$R < 16$\,kpc at \wave{3.6} agree with the extensions of the lines fitted
through the points at the lower frequencies. This indicates that the base level
of the map at \wave{3.6} outside the central area is correct. Therefore, we
can calculate spectral index maps between \wave{20.5} \citep{beck98}
and \wave{3.6} of total emission ($\alpha$) and non-thermal emission ($\alphan$)
at the best available angular resolution of 90\arcsec\ HPBW.

Before calculating spectral index maps, we subtracted from both maps the same
point sources (i.e. all sources with flux densities above 5\,mJy
at \wave{20.5} and above 1.2\,mJy at \wave{3.6}).
In order to obtain maps of non-thermal emission at these frequencies,
we subtracted maps of thermal emission from the maps of total emission.
%$NTH = I - TH$
We used the thermal map at \wave{20.5} that
\citet{taba13a} derived from the extinction-corrected H$\alpha$ map of
\citet{devereux94}, which we smoothed to 90\arcsec\ HPBW and scaled  to
\wave{3.6} by the factor $(3.6/20.5)^{0.1}$, using the thermal spectral index of
0.1. Both $I$ maps were cut down to the size of the thermal map of
$110\arcmin \times 39\arcmin$ in $l \times b$ before subtracting $TH$ from $I$, and
all maps were transformed onto the same grid. Figure~2 in \citet{taba13a} shows
that due to the propagation of cosmic ray electrons (CREs) the distribution
of the $NTH$ emission is much more extended than that of the $TH$ emission.

For the spectral index calculation, only data points above two times the
noise level in both maps were used. The resulting maps of $\alpha$ and
$\alphan$ are shown in Figures~\ref{fig:alpha} and \ref{fig:alpha_nth},
respectively. We note that because of the base level
problems in the innermost region   $10\arcmin \times 10\arcmin$ in size at
\wave{3.6}, the spectral indices cannot be trusted there.

In Figure~\ref{fig:alpha}, values of $\alpha$ in the ring vary from
about 0.4 in the middle of the ring where the $\HII$ regions are located
to $> 1.0$ in the outer regions of the ring where most of the emission is
$NTH$. The values of $\alphan$ in Figure~\ref{fig:alpha_nth} show the
same trend as those of $\alpha$, but are about 0.1 higher in the $\HII$ regions
in the middle of the ring, varying between 0.5 here and $> 1.0$ in the outer
regions. A value of $\alphan = 0.5$ near star-forming regions indicates
that the CREs are still close to their birth places in the supernova
remnants. Energy losses during the propagation away from their birth places
cause the larger spectral indices in the outer parts of the ring.

In Figures~\ref{fig:alpha} and \ref{fig:alpha_nth}
the spectral indices are larger in the southern
half of M~31 (right-hand part) than in the northern half. The $\alpha$
map of \citep{berkhuijsen03} between \wave{20.5} and \wave{6.2}
shows the same trends as seen in Figure~\ref{fig:alpha}, but with less
detail because of the larger HPBW of $3\arcmin$.

%%----------------------------------------
%FIG15 Radial spectral index distributions
\begin{figure}[htbp]
\begin{center}
\includegraphics[width=0.8\columnwidth,angle=270]{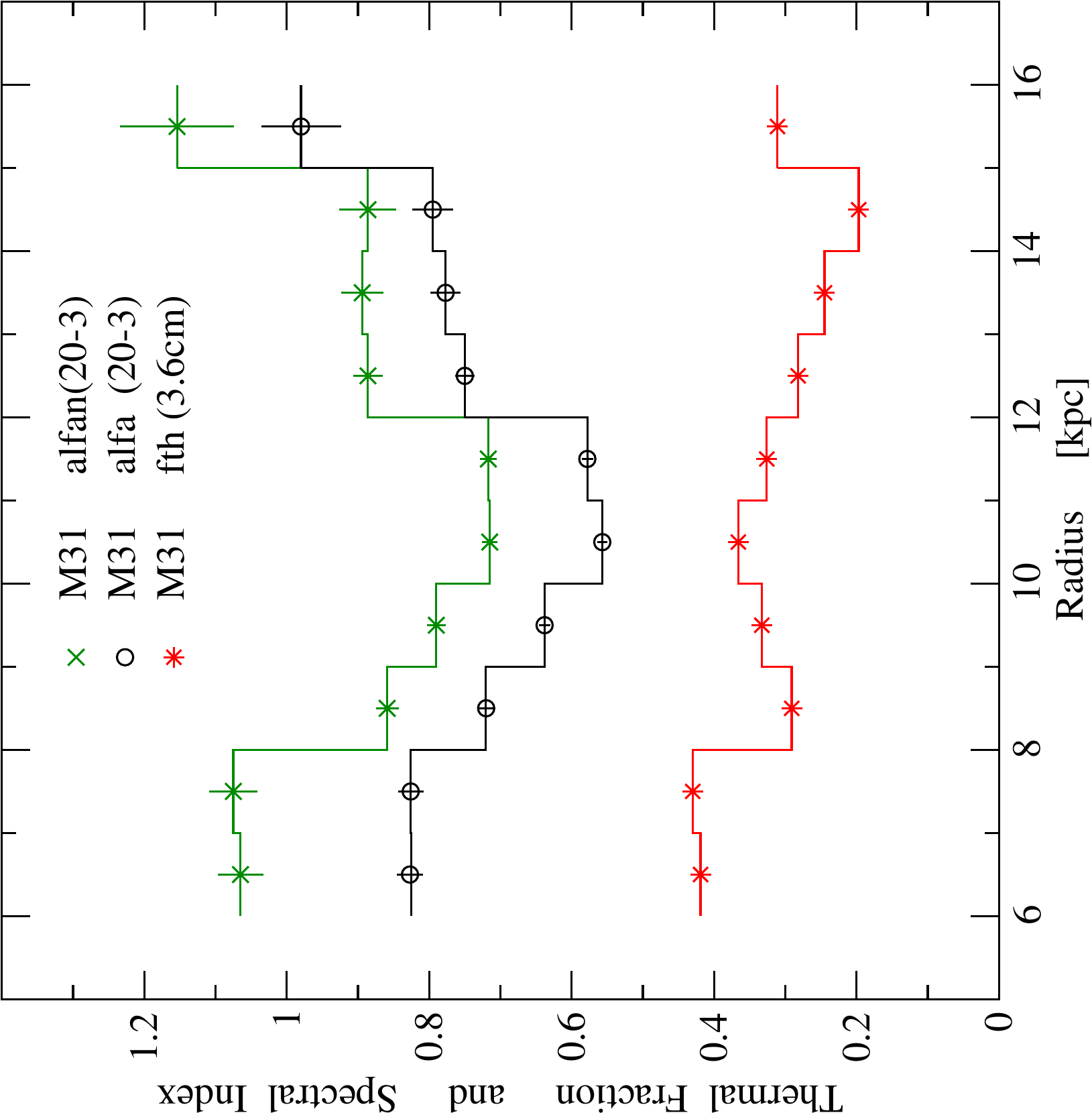}
\hfill
\caption{Radial variations of the spectral index between \wave{20.5} and
  \wave{3.6}, of the total intensity (black),  of the non-thermal intensity
  (green), and of the thermal fraction at \wave{3.6} (red); they are all at
  $1\farcm5$ HPBW.}
\label{fig:alpha_radial}
\end{center}
\end{figure}
%%----------------------------------------

Due to the large frequency interval between \wave{20.5} and \wave{3.6},
the random noise errors in the spectral index maps are quite small,
ranging from 0.013 in the middle of the ring to 0.038 in the outer
parts of the ring in $\alpha$ (Figure~\ref{fig:alpha}) and from 0.023
to 0.046 in $\alphan$ (Figure~\ref{fig:alpha_nth}). The errors are
dominated by the noise errors in the maps at \wave{3.6}. Systematic errors
due to base level uncertainties are difficult to estimate, but could be
larger than the random noise errors.

We also calculated spectral index maps between \wave{20.5} and \wave{6.2}
and between \wave{6.2} and \wave{3.6} at the resolution of $3\arcmin$ HPBW.
They are consistent with Figures~\ref{fig:alpha} and \ref{fig:alpha_nth},
but because of the larger beam width and the narrower frequency ranges,
less interesting than the higher resolution spectral index maps between
\wave{20.5} and \wave{3.6} shown here. The $NTH$ map at \wave{6.2} is
shown in Fig.~\ref{fig:cm6nth}.

%-----------------------Subsection    4.3----------------------------

\subsection{Radial variations in spectral index}
\label{subsect:radspec}

In Figure~\ref{fig:alpha_radial} we show the radial variation in $\alpha$ and
$\alphan$, and the fraction of thermal emission ($f_{th}$), averaged in 1 kpc
wide rings between $R = 6$\,kpc and $R = 16$\,kpc. As discussed above,
the flattest spectra occur in the middle of the ring ($R = 10-11$\,kpc)
where most of the $\HII$ regions are located and the $TH$ fraction of the
emission is highest. Due to the subtraction of the thermal emission,
the spectrum of the $NTH$ emission is typically
steeper by about 0.1 than that of the total emission $I$, and both spectra become
significantly steeper towards the edges of the emission ring where the $NTH$
emission dominates. This suggests that CREs move  inwards and
outwards away from their birth
places near star-forming regions over several kpc in radius.

%----------------------------SECTION   5--------------------------

\section{Radial scale lengths}
\label{sec:scale}

It is interesting to see how the various types of emission from M~31
vary with radial distance to the galaxy centre.

\begin{figure}[htbp]
\begin{center}
\includegraphics[width=0.8\columnwidth,angle=270]{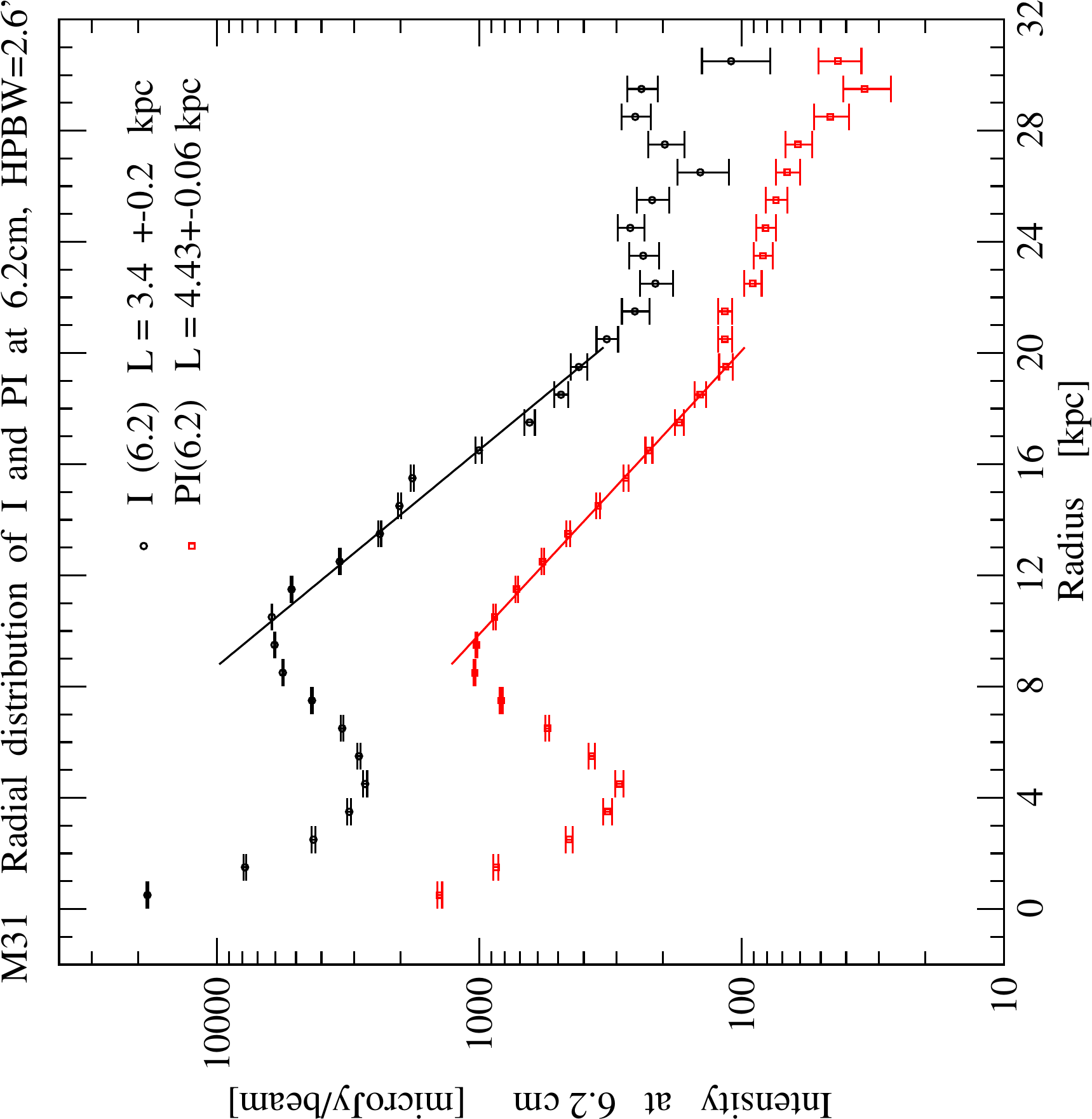}
\hfill
\caption{Radial variations of total intensity $I$ (black) and polarized
  intensity $PI$ (red) at \wave{6.2} at $2\farcm6$ HPBW, derived from
  the full extent of the observed area. This map is complete out
  to $R = 18$\,kpc (or $80\arcmin$) along the minor axis. The exponential
  fits were restricted to the radial range between 9\,kpc and 20\,kpc.
  The scale lengths $L$ are given in the figure.}
\label{fig:radial}
\end{center}
\end{figure}

In Figure~\ref{fig:radial} we show the radial distributions of
total emission and
polarized emission in the new map at \wave{6.2} with angular
resolution of $2\farcm6$ HPBW.
The emissions were averaged in 1 kpc wide circular
rings in the plane of M~31 using a constant inclination of $75\degr$. As
this map extends $140\arcmin$ in longitude, the radial distribution is
complete out to $R = 18$\,kpc, beyond which some emission near the major
axis is missing.   The $I$ emission from the bright ring peaks at
$R = 10-11$\,kpc and the $PI$  emission at 9\,kpc, then  both emissions
steadily decrease to about $R = 20$\,kpc. Exponential fits for the range
$R = 9 - 20$\,kpc yield radial scale lengths of $L = (3.4 \pm 0.2)$ kpc
in $I$ and $L = (4.43 \pm 0.06)$\,kpc in $PI$, as indicated in the figure.

%%----------------------------------------
%FIG16 Radial distributions I, PI
%%----------------------------------------

%%----------------------------------------
%FIG17 Radial distributions NTH, TH, PI
\begin{figure}[htbp]
\begin{center}
\includegraphics[width=0.8\columnwidth,angle=270]{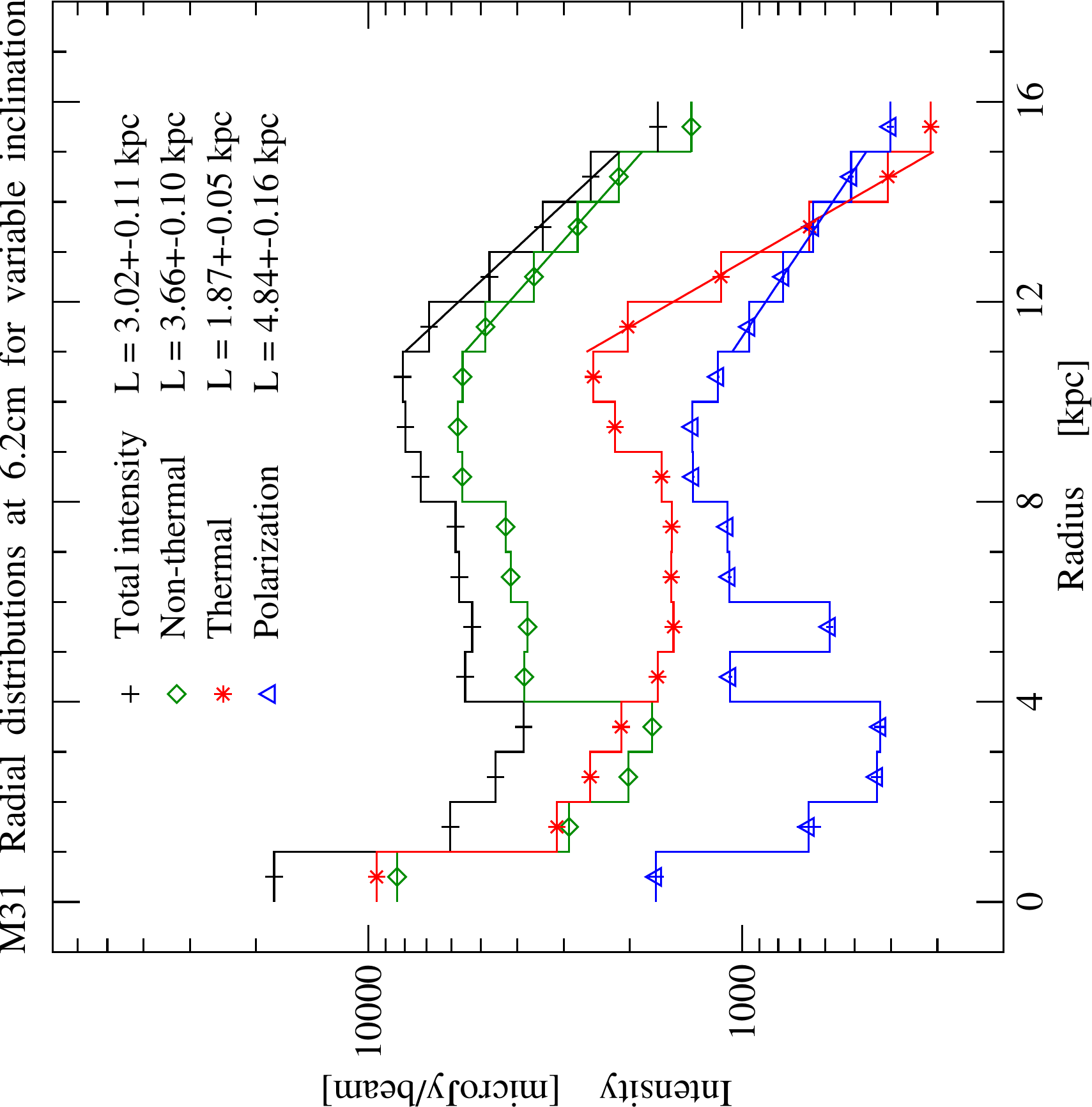}
\hfill
\caption{Radial variations of total intensity $I$ (black), non-thermal
    intensity $NTH$ (green), thermal intensity $TH$ (red), and polarized
    intensity $PI$ (blue)
    at \wave{6.2} at $3\arcmin$ HPBW. The exponential fits were
    restricted to the radial range between 11\,kpc and 15\,kpc.
    The scale lengths $L$ are given in the figure.}
\label{fig:radial2}
\end{center}
\end{figure}
%%----------------------------------------

%---------------------------------------------------------------------------
%TABLE4
\begin{table*}
\begin{center}
  \caption{Exponential scale lengths $L$ for the radial range 11--15\,kpc
    at three frequencies for total intensity $I$, non-thermal intensity
    $NTH$, polarized intensity $PI$, and thermal intensity $TH$.}
  \label{tab:scale}
  \begin{tabular}{llccc}
  \hline
  Resolution & Component & L (\wave{20.5}) & L (\wave{6.2}) & L (\wave{3.6}) \\
  $[\arcmin]$&           & [kpc]         & [kpc]         & [kpc] \\
  \hline
  3          & I  & $3.66\pm0.05$   &   $3.02\pm0.11$  &   $2.46\pm0.14$ \\
  3          & NTH & $4.08\pm0.13$   &   $3.66\pm0.10$  &   $2.79\pm0.06$ \\
  3          & PI  &  --             &   $4.84\pm0.16$  &   $3.79\pm0.18$ \\
  3          & TH  & $1.90\pm0.05$   &   $1.87\pm0.05$  &   $1.87\pm0.05$ \\
  \hline
  1.5        & I  & $3.26\pm0.14$   &     --           &   $2.35\pm0.34$ \\
  1.5        & NTH & $3.60\pm0.09$   &     --           &   $2.72\pm0.44$ \\
  1.5        & PI  &  --             &     --           &   $3.00\pm0.25$ \\
  1.5        & TH  & $1.69\pm0.15$   &     --           &   $1.69\pm0.15$ \\
  \hline
  \end{tabular}
\end{center}
\end{table*}
\normalsize
%---------------------------------------------------------------------------

The radial variations of $I$, $NTH$, $TH$, and $PI$ at \wave{6.2}
at $3\arcmin$ HPBW are shown in Figure~\ref{fig:radial2}. For the
range $R = 0 - 7$\,kpc, we used the inclinations of the $\HI$ gas
increasing from $31\degr$ at $R < 2$\,kpc to $72\degr$ at $R = 6-7$\,kpc as
determined by \citet{chemin09}, assuming that the same holds for the radio
continuum emission. A close correspondence between gas and radio continuum
features was first pointed out by \citet{beck82} and \citet{berkhuijsen93}
and is clearly visible in Figure~8 of \citet{nieten06}. At larger radii we
used the inclination of $75\degr$ again, which is consistent with the
inclination of $\HI$ in this area. For each component we calculated the
radial scale length between $R = 11$\,kpc and $R = 15$\,kpc, as shown in
Figure~\ref{fig:radial2} and listed in Table~\ref{tab:scale}.
With $L = (3.66 \pm 0.10)$\,kpc
for $NTH$ and $L = (1.87 \pm 0.05)$\,kpc for $TH$, the $NTH$ emission
clearly decreases much more slowly than the $TH$ emission, indicating
propagation of the CREs away from their birth places in the star-forming regions.
The scale length of $PI$ of $L = (4.84 \pm 0.16)$\,kpc is even larger
than that of $NTH$, reflecting the large scale of the ordered magnetic
field without influence of the turbulent fields of smaller scale that
dominate the $NTH$ emission.

Since propagation of CREs depends on frequency, we also calculated
the scale lengths at \wave{20.5} and \wave{3.6} at the angular
resolution of $3\arcmin$ and at the resolution of $1\farcm5$. As
before, we determined the scale lengths between 11\,kpc and 15\,kpc. All
scale length results are given in Table~\ref{tab:scale}.

Comparing the scale lengths at the resolution of $3\arcmin$, we see a
clear decrease in $L$ of $I$ and $NTH$ with increasing frequency: $L$
for $NTH$ drops from $(4.08 \pm 0.13)$\,kpc at \wave{20.5} to
$(2.79 \pm 0.06)$\,kpc at \wave{3.6}. This reflects the decrease in the
propagation length of the CREs with increasing frequency and makes $NTH$ maps
at \wave{20.5} look smoother than those at \wave{3.6}. In addition, the scale
length of $PI$ decreases between \wave{6.2} and \wave{3.6}, but not as
strongly as that of $NTH$, because $PI$ depends on the large-scale ordered
magnetic field and $NTH$ on both the ordered and the small-scale
turbulent field. Naturally, the scale length of $TH$ is the same at
each frequency.

At the resolution of $1\farcm5$ the scale lengths of $I$ and $NTH$ at
\wave{20.5} are again larger than those at \wave{3.6} by nearly the
same amount as at $3\arcmin$ resolution. However,  the scale lengths of $I$
and $NTH$ at $1\farcm5$ are significantly smaller than those at $3\arcmin$.
Although at \wave{3.6} the errors are quite large, the same trend is visible
and is  clearest for the scale length of $PI$. The
scale lengths are larger at $3\arcmin$ resolution  than at $1\farcm5$
resolution, because of the greater smoothing of the emission at $3\arcmin$
resolution than at $1\farcm5$ resolution. At \wave{3.6} the effect is
smaller than at \wave{20.5} because the propagation length at the higher frequency
is smaller than at \wave{20.5}. As $TH$ emission is least diffuse, at both
frequencies the scale length of $TH$ is only slightly smaller at $1\farcm5$
than at $3\arcmin$.

%------------------------------SECTION--------6---------------------------

\section{Azimuthal variation of polarized intensity}
\label{sec:PI}

%%----------------------------------------
%FIG18 Azimuthal variation of I(unpol) and PI at 3cm
\begin{figure}[htbp]
\begin{center}
\includegraphics[width=0.7\columnwidth,angle=270]{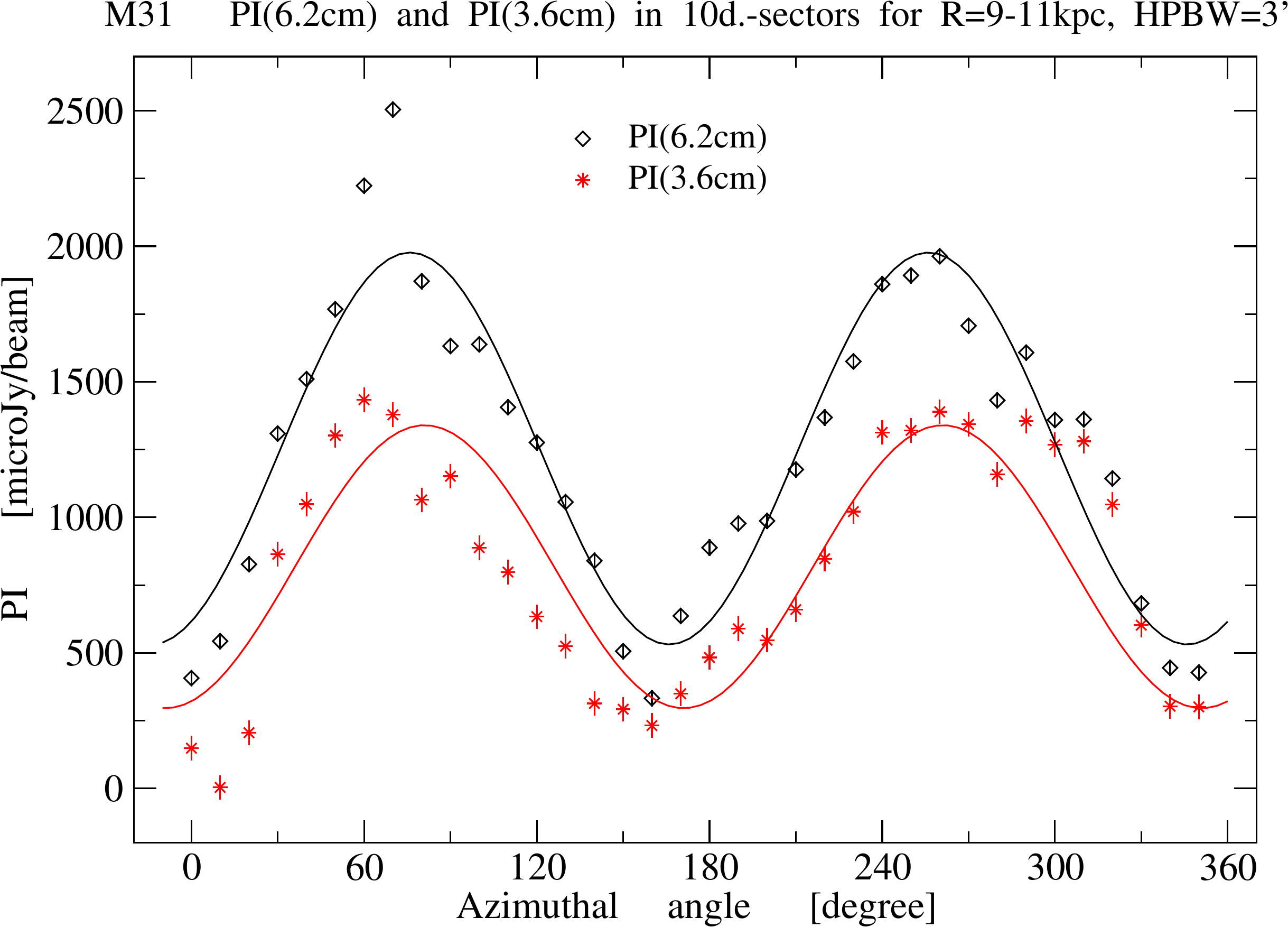}
\hfill
\caption{Variation of polarized intensity $PI$ at \wave{3.6} (red)
and \wave{6.2} (black) at
$3\arcmin$ HPBW with azimuthal angle $\phi$ in the plane of M~31,
  averaged in $10\degr$   sectors in the ring between 9\,kpc and
  11\,kpc radius. The azimuthal angle is counted anticlockwise from
  the north-eastern major axis of the ring in the plane of the sky
  (see Fig.~\ref{fig:sectors}). The lines show the weighted fits.}
\label{fig:PI_azm}
\end{center}
\end{figure}
%%----------------------------------------

%%----------------------------------------
%FIG19 Sectors and PI at 6cm
\begin{figure}[htbp]
\begin{center}
\includegraphics[width=0.9\columnwidth]{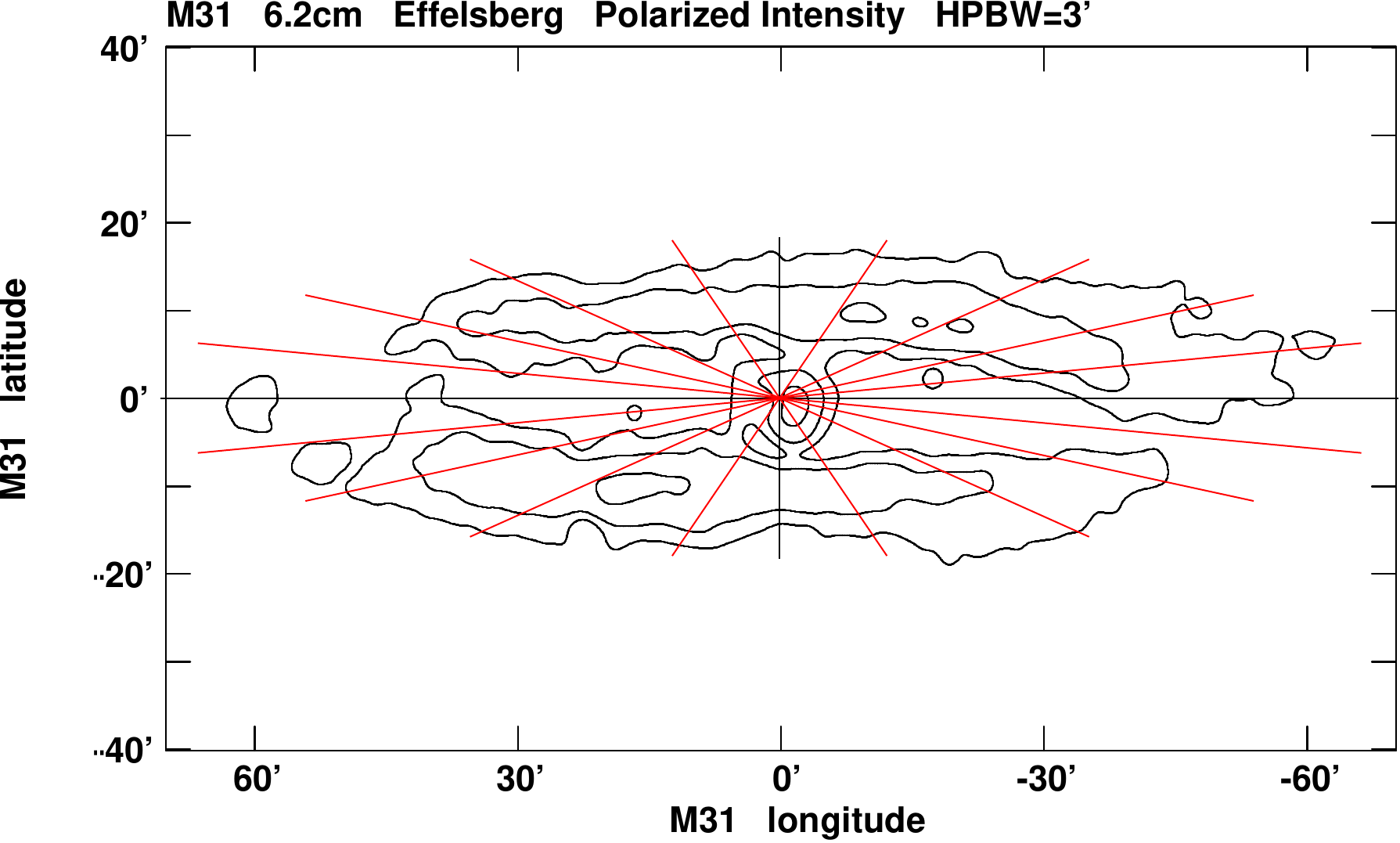}
\hfill
\caption{Sectors of $20\degr$ in width in the galaxy plane (red lines)
and major and minor axes (black lines), superimposed onto contours of
polarized intensity $PI$ at \wave{6.2} at $3\arcmin$ HPBW.
Contour levels are at (0.6, 1.2, 2.4) $\times$ 1\,mJy/beam.
The azimuthal angle is counted anticlockwise from
the north-eastern major axis of the ring (left side).
%The HPBW is indicated in the bottom left corner.
}
\label{fig:sectors}
\end{center}
\end{figure}
%%----------------------------------------

%%----------------------------------------
%FIG20 Azimuthal variation of PI
\begin{figure}[htbp]
\begin{center}
\includegraphics[width=0.7\columnwidth,angle=270]{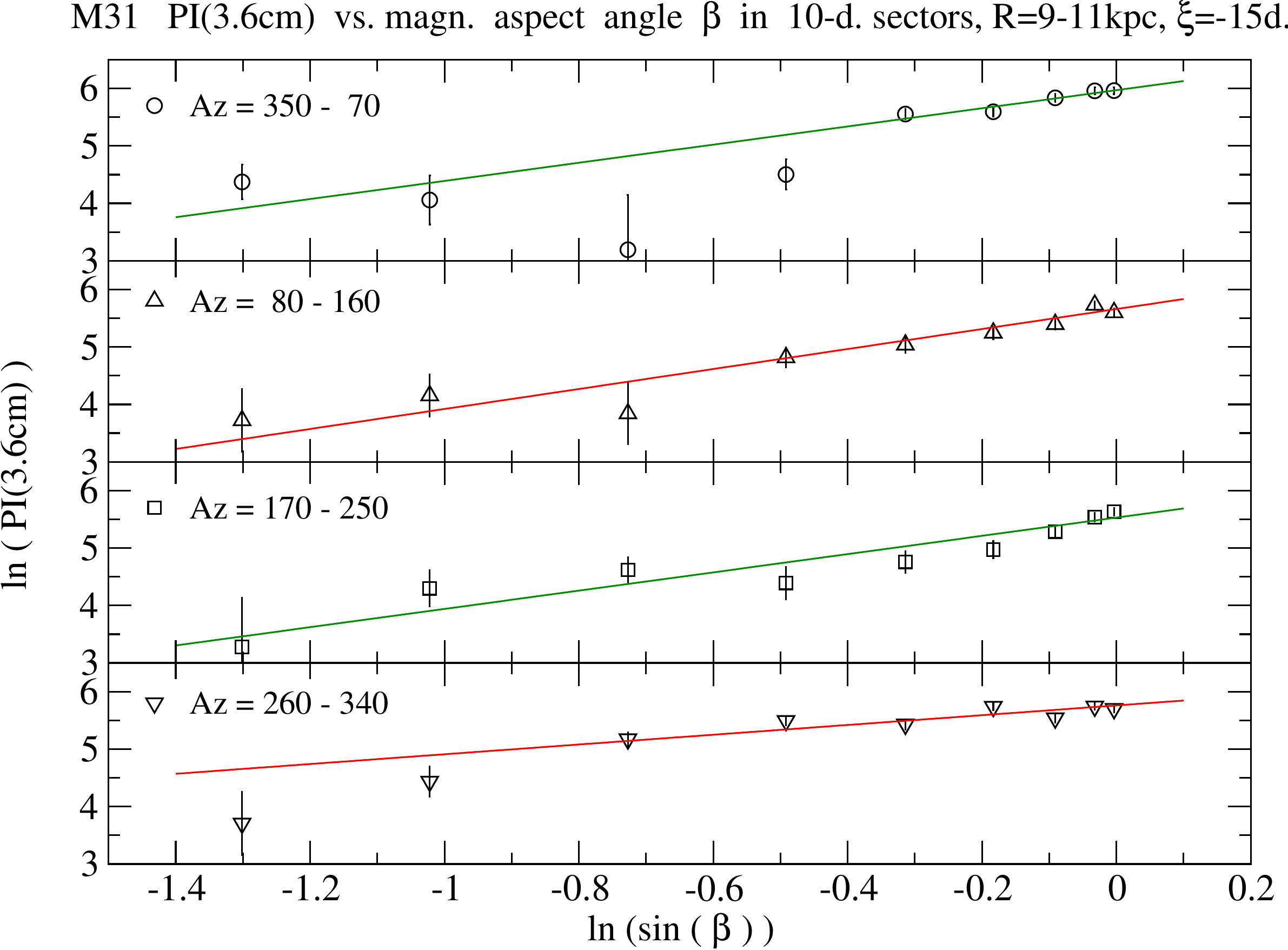}
\hfill
\caption{Variation of polarized intensity $PI$ at \wave{3.6}
(in $\mu$Jy/beam, in log$_\mathrm{e}$ scale)
at $1\farcm5$ HPBW in four quadrants of M~31 with aspect angle
$sin\,\beta_{\mathrm{ord}}$ (in log$_\mathrm{e}$ scale)
of the line of sight towards a spiral field with a pitch angle of
$\xi_\mathrm{ord}=-15\degr$ for the radial ring between 9\,kpc and
11\,kpc. The lines show weighted fits through the data points.
The fit results are given in Table~\ref{tab:quadrants}.
}
\label{fig:PI_quadrants}
\end{center}
\end{figure}
%%----------------------------------------

Polarized intensity $PI$ is proportional to the component of the ordered
(i.e. regular plus anisotropic turbulent) field perpendicular to
the line of sight, $B_{\mathrm{ord},\perp}$.
The strength of the $PI$ depends on the strength and geometry of the ordered field,
the density of the CREs, and the amount of depolarization.

Figure~\ref{fig:PI_azm} shows the azimuthal variation of polarized
intensity at \wave{3.6} and \wave{6.2} in the emission ring.
Polarized intensity reveals maxima near the minor axis and minima
near the major axis.
%The similarity of the two curves shows that wavelength-dependent
%Faraday depolarization is small at these wavelengths.
Fits of a double-periodic curve give phase shifts of
$-9\degr\pm3\degr$ at \wave{3.6} and $-14\degr\pm2\degr$ at \wave{6.2},
with the latter fit being statistically better.
%Unpolarized intensity is emitted by CREs in unresolved (isotropic turbulent or tangled) magnetic fields in the sky plane, while polarized intensity emerges
%from ordered fields in the sky plane. Hence, Fig.~\ref{???} shows that unresolved fields are spread over the ring of strongest star formation.
Variations in the CRE density or the strength of the ordered field
are independent of the locations of the major and minor axis
and cannot explain the variation seen in Fig.~\ref{fig:PI_azm}.
Depolarization by Faraday dispersion in turbulent magnetic fields
could in principle increase from the minor to the major axis
due to the increasing path length  through the emission ring. However,
the emission ring between 9\,kpc and 11\,kpc radius is flat, with a
scale height of the thermal gas of only about 0.5\,kpc \citep{fletcher04}
compared to a radial width of about 4\,kpc (Fig.~\ref{fig:radial2}),
resulting in an almost constant path length between the minor and the major
axis, which disfavours a strong variation in Faraday depolarization.
Depolarization by $RM$ gradients \citep{fletcher04} is strongest near
the minor axis (Fig.~\ref{fig:rm3_6_azm}) and also cannot explain
the variation in Fig.~\ref{fig:PI_azm}.
Furthermore, Faraday depolarization is strongly wavelength dependent and
is expected to be weak at \wave{3.6} and \wave{6.2}.
In the next paper, we will discuss Faraday depolarization in detail.

We conclude that the variation in polarized intensity is due to
  %(a) variations in CRE density, (b) variations in the strength of the
  %ordered field, or (c)}
variation in the orientation of the ordered field with respect to the
line of sight. The  polarized intensity is strongest around
the minor axis and weakest around the major axis, which indicates that
%c) is the dominant effect and
the orientation of the ordered field approximately follows the ring,
as is clearly visible in Figs.~\ref{fig:cm11pi2} and \ref{fig:cm6nth},
so that its component in the plane of the sky varies with azimuthal angle
with a phase shift that is related to the spiral pitch angle
$\xi_\mathrm{ord}$ of the large-scale ordered field.

To investigate this geometrical effect in detail, we computed the
aspect angle $\beta_{\mathrm{ord}}$ between the ordered magnetic field and the
line of sight, so that $B_{\mathrm{ord},\perp}=B_{\mathrm{ord}} \,sin\,\beta_{\mathrm{ord}}$
and $B_{\mathrm{ord},\parallel}=B_{\mathrm{ord}} \,cos\,\beta_{\mathrm{ord}}$.
For a large-scale $ASS$ pattern of the ordered field
with a constant pitch angle $\xi_\mathrm{ord}$,
\begin{equation}
\label{eq:aspect}
cos\,\beta_{\mathrm{ord}} = cos\,(\phi-\xi_\mathrm{ord}) ~ sin\,i\, ,
\end{equation}
where $i=75\degr$ is the galaxy inclination, and $\phi$ is the azimuthal
angle in the galaxy plane, counted anticlockwise,
with $\phi=0\degr$ on the north-eastern and $\phi=180\degr$ on the
south-western major axis of the ring
(see Fig.~\ref{fig:sectors}). For a constant CRE
density, constant strength of the ordered field, constant  path length
through the emitting ring, and negligible Faraday depolarization,
polarized intensity $PI$ varies as
\begin{equation}
\label{eq:PI}
PI ~ \propto ~ N_\mathrm{CRE} ~ B_{\mathrm{ord},\perp}^{\,\,\,\,1\,+\,\alpha_\mathrm{nth}} ~ \propto ~ |~(sin\,\beta_{\mathrm{ord}})^{\,\,1\,+\,\alpha_\mathrm{nth}}~| \, ,
\end{equation}
where $\alpha_\mathrm{nth}$ is the synchrotron spectral
index.\footnote{Eq.~\ref{eq:PI} is also valid for the case of equipartition
  between the energy densities of cosmic rays and total magnetic fields
  ($N_\mathrm{CRE} \propto N_\mathrm{CR} \propto B_\mathrm{tot}^2$) because
  $B_\mathrm{tot}$ is dominated by small-scale turbulent fields that do
   not depend on $\beta_{\mathrm{ord}}$.}
According to Eq.~\ref{eq:PI}, the maxima of $PI$ are expected at
$\beta_{\mathrm{ord}}=90\degr$ (i.e. at azimuthal angles of
$\phi=(90\degr+\xi_\mathrm{ord})$ and $\phi=(270\degr+\xi_\mathrm{ord})$), and
the minima at $\beta_{\mathrm{ord}}=(90\degr - i)$ (i.e. at
$\phi=+\xi_\mathrm{ord}$ and $\phi=(180\degr+\xi_\mathrm{ord})$).

%---------------------------------------------------------------------------
%TABLE5
\begin{table}
\begin{center}
  \caption{Results of the fitted lines shown in
  Fig.~\ref{fig:PI_quadrants} for the four quadrants of the radial ring between
  9\,kpc and 11\,kpc. The first column gives the range of azimuthal angle,
  the second the slope and its error, and the third the $\chi^2$ value.}
  \label{tab:quadrants}
  \begin{tabular}{ccc}
  \hline
  Azimuth [$\degr$] & Slope & $\chi^2$ \\
%  [$\degr$]\\
  \hline
  $350-70$   & $1.58\pm0.25$ & 1.8 \\
  $80-160$   & $1.74\pm0.22$ & 1.0 \\
  $170-250$  & $1.59\pm0.27$ & 1.6 \\
  $260-340$  & $0.85\pm0.20$ & 2.0 \\
  \hline
  \end{tabular}
\end{center}
\end{table}
\normalsize
%---------------------------------------------------------------------------

When plotting $log_\mathrm{e}(PI)$ against $log_\mathrm{e}(sin~\beta_{\mathrm{ord}})$,
the slope of the fit should give 1\,+\,$\alpha_\mathrm{nth}$ in the case of a
perfect $ASS$ field and no depolarization. Table~\ref{tab:quadrants} shows the results
for four quadrants of the radial ring between 9\,kpc and 11\,kpc,
each covering a range of azimuthal angles corresponding to a range
in aspect angle between $\beta_{\mathrm{ord}}=(90\degr - i)$ and
$\beta_{\mathrm{ord}}=90\degr$.
Here we assumed that the ordered field has a large-scale  $ASS$ pattern with a pitch angle of
%$\xi_\mathrm{ord}=-26\degr$,
%taken from the average pitch angle determined from Fig.~\ref{fig:pw3_6_azm}.
$\xi_\mathrm{ord}=-15\degr$.\footnote{While the fit to the \wave{6.2}
data in Fig.~\ref{fig:PI_azm} gave
$\xi_\mathrm{ord}=-14\degr\pm2\degr$, we decided to use a value of
$-15\degr$, which gives a symmetric assignment of the azimuthal sectors to the
four quadrants.}
The slopes of the first three quadrants are consistent with the mean
synchrotron spectral index of about 0.75 (Fig.~\ref{fig:alpha_radial}).
In these quadrants, the variation of $PI$ is mostly due to the variation
in aspect angle $\beta$. In the fourth quadrant (north)
($\phi=260\degr-340\degr$) the slope is significantly different from the
expectation for the simple geometry. In this region the ordered field deviates
from the assumed $ASS$ geometry in strength and/or in orientation.
%The possible effects of Faraday depolarization will be discussed in the next paper.

%%----------------------------------------
%FIG21 RM(6/11cm)
\begin{figure*}[htbp]
\begin{center}
\includegraphics[width=10cm]{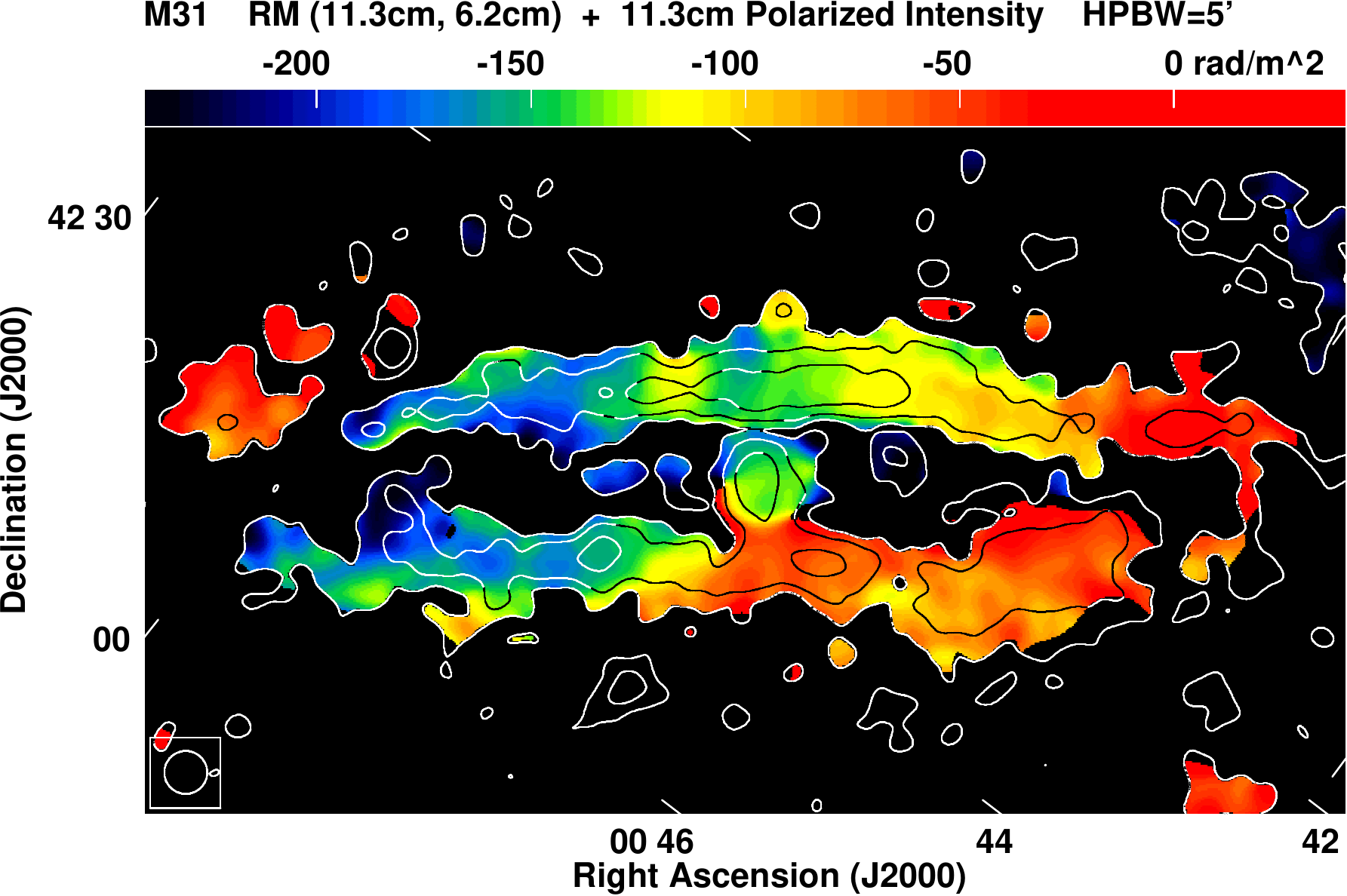}
\hfill
\caption{Faraday rotation measure $RM$ (colour) between \wave{11.3} and
  \wave{6.2} at $5\arcmin$ HPBW, calculated at pixels where $PI$ at
  both frequencies exceeds three times the rms noise. The error decreases
  from $26$\,rad\,/m$^2$ at the lowest signal-to-noise ratio ($S/N=3$) to
  about $3$\,rad\,/m$^2$ at the highest $S/N$. Contours show the polarized
  intensity at \wave{11.3} at the same resolution. Contour levels are at
  1, 2, and 4\,mJy/beam. Polarized background sources have been subtracted.
  The HPBW is indicated in the bottom left corner.
  The coordinate system is rotated by $-53\degr$.
  }
\label{fig:rm6_11}
\end{center}
\end{figure*}
%%----------------------------------------

%%----------------------------------------
%FIG22 RM(3/6cm)
\begin{figure*}[htbp]
\begin{center}
\includegraphics[width=10cm]{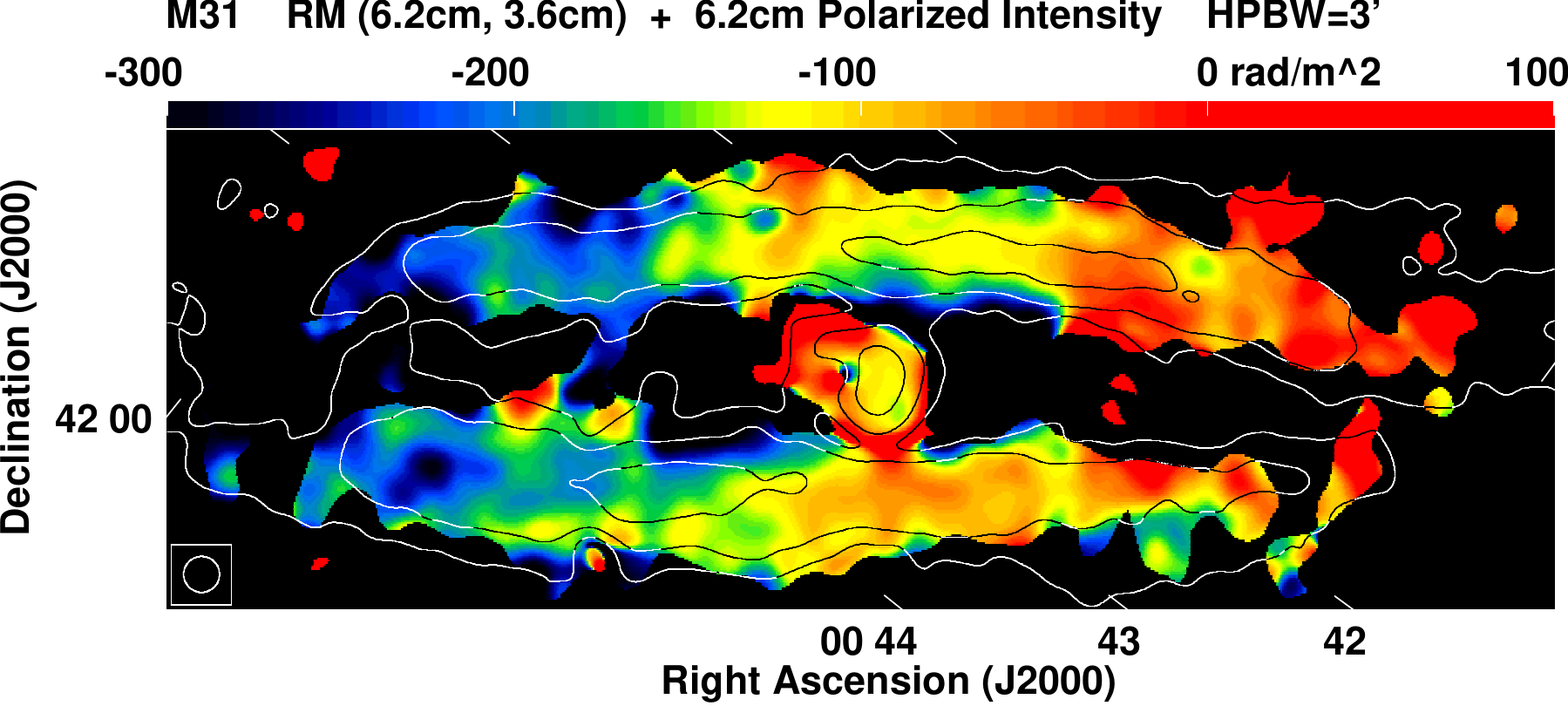}
\hfill
\caption{Faraday rotation measure $RM$ (colour) between \wave{6.2} and
  \wave{3.6} at $3\arcmin$ HPBW, calculated at pixels where $PI$ at
  both frequencies exceeds three times the rms noise. The error decreases
  from $93$\,rad\,/m$^2$ at $S/N=3$ to about $9$\,rad\,/m$^2$ at the
  highest $S/N$. Contours show the polarized intensity at \wave{6.2} at
  the same resolution. Contour levels are at 0.5, 1, and 2\,mJy/beam.
  Polarized background sources have been subtracted.
  The HPBW is indicated in the bottom left corner.
  The coordinate system is rotated by $-53\degr$.
  }
\label{fig:rm3_6}
\end{center}
\end{figure*}
%%----------------------------------------

In a similar study by \citet{beck82} of $PI$ at \wave{11.1} (Fig.~6 therein),
the slopes were different from the expected value in all four quadrants,
probably due to the lower resolution and significant
Faraday depolarisation at that wavelength. Our new results are more
consistent with the $ASS$ field pattern.

\citet{beck82} suggested that
the field orientation follows gaseous spiral arms observed in $\HI$
\citep{unwin80a,unwin80b,braun90,chemin09} that deviate from a simple ring
structure. The main $\HI$ spiral arm in the north-western quadrant is almost
straight from the minor axis to close to the major axis, so that $\beta$
hardly varies, which may explain the small slope in
Fig.~\ref{fig:PI_quadrants} (bottom panel).

%---------------------SECTION------7-----------------------------------

\section{Faraday rotation and large-scale field pattern}
\label{sec:rm}

%%----------------------------------------
%FIG23 Azimuthal variation of RM(6/11cm)
\begin{figure}[htbp]
\begin{center}
\includegraphics[width=0.6\columnwidth,angle=270]{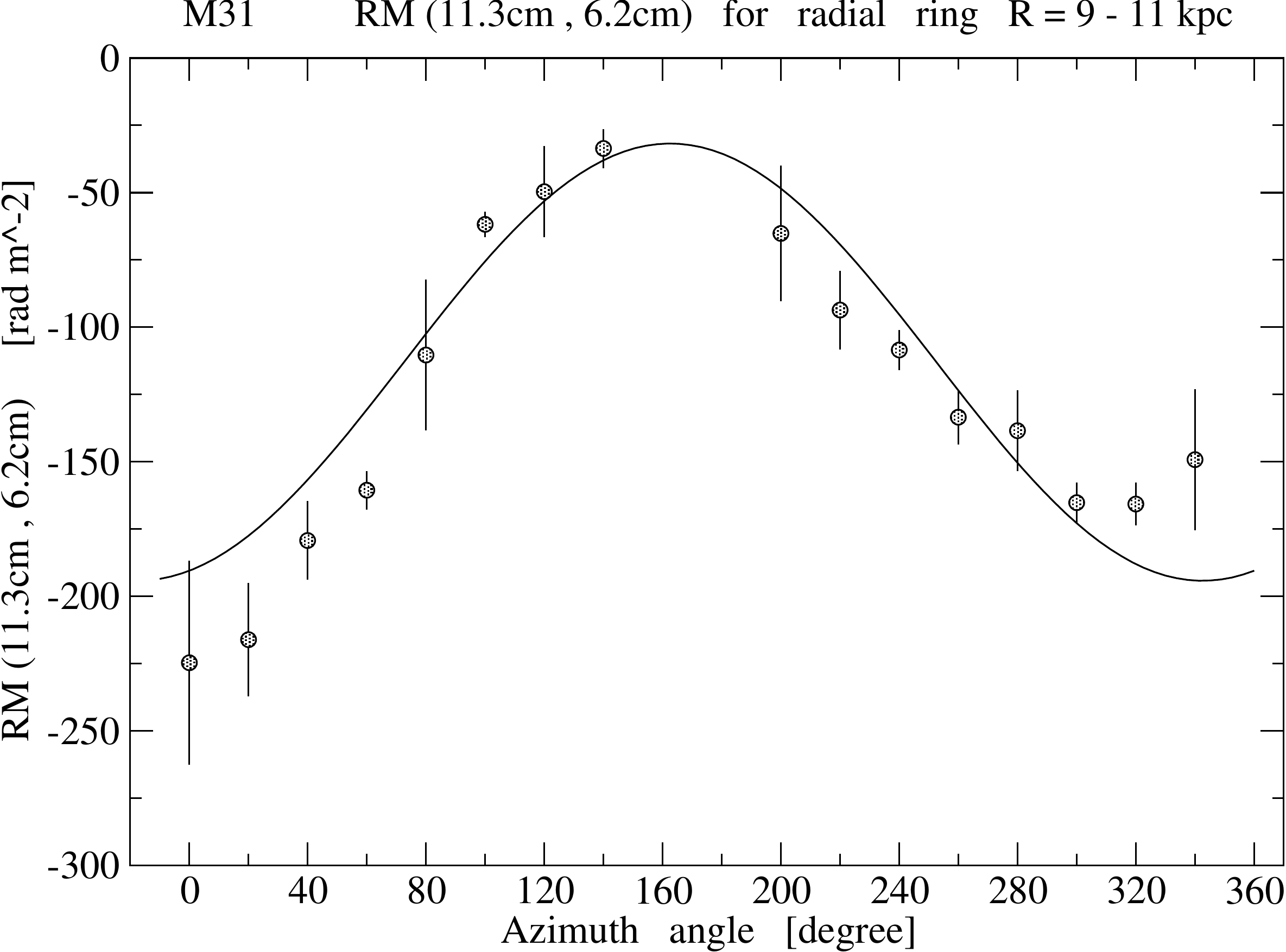}
\hfill
\caption{Variation (with azimuthal angle in the galaxy plane) of Faraday
  rotation measures $RM$ between \wave{11.3} and \wave{6.2} at $5\arcmin$
  HPBW, averaged in the radial ring between 9\,kpc and 11\,kpc
  in sectors of $20\degr$ width,
  and the fitted sinusoidal line. The azimuthal angle is counted from
  the north-eastern major axis (left side in Fig.~\ref{fig:rm6_11}).
  The reduced $\chi^2$ value is 4.0.
  }
\label{fig:rm6_11_azm}
\end{center}
\end{figure}
%%----------------------------------------

%%----------------------------------------
%FIG24 Azimuthal variation of RM(3/6cm)
\begin{figure}[htbp]
\begin{center}
\includegraphics[width=0.6\columnwidth,angle=270]{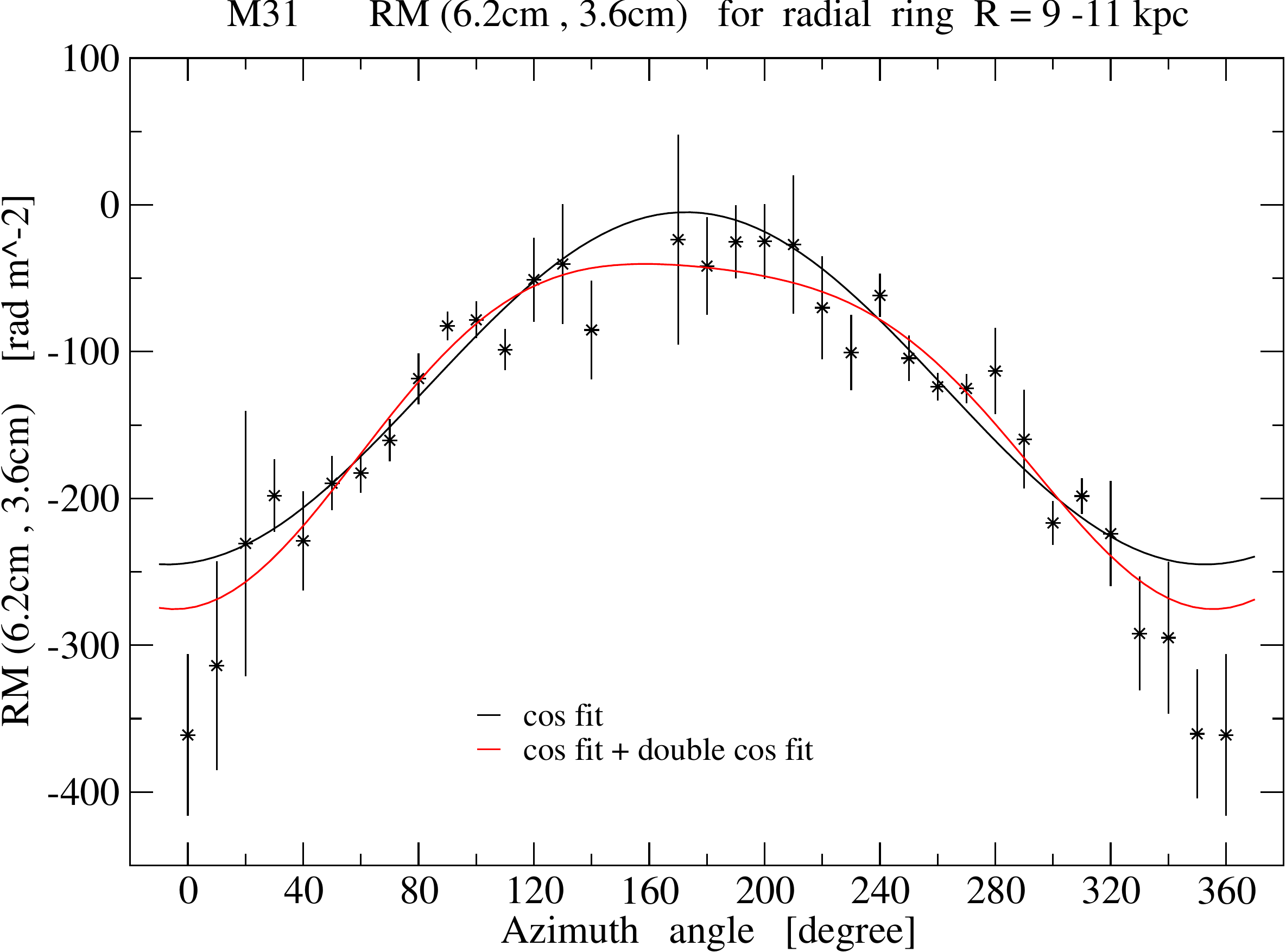}
\hfill
\caption{Variation (with azimuthal angle in the galaxy plane) of Faraday
  rotation measures $RM$ between \wave{6.2} and \wave{3.6} at $3\arcmin$
  HPBW, averaged in the radial ring between 9\,kpc and 11\,kpc.
  in sectors of $10\degr$ width. The black line shows
  the sinusoidal fit, the red line the combined sinusoidal +
  double-periodic fit, with reduced $\chi^2$ values of 1.3 for both fits.
%The azimuthal angle is counted from the north-eastern major axis (left side in Fig.~\ref{fig:rm3_6}).
}
\label{fig:rm3_6_azm}
\end{center}
\end{figure}
%%----------------------------------------

%---------------------------------------------------------------------------
%TABLE6
\begin{table*}
\begin{center}
  \caption{Fit results of the sinusoidal azimuthal variation
  of $RM$ between \wave{6.2} and \wave{3.6} at $3\arcmin$ HPBW,
  averaged in sectors of $10\degr$ width,
  in five radial rings $\Delta R$ in the galaxy plane:
  foreground $RM_\mathrm{fg}$; amplitude $RM_\mathrm{max,a}$; phase,
  corresponding to the average pitch angle $\xi_\mathrm{reg}$ of the regular
  field in the radial ring, and reduced $\chi^2$ of the fit.
  The last column gives the average pitch angle $\xi_\mathrm{ord}$ of the
  ordered field, calculated from the $B$ orientations, corrected for Faraday
  rotation, and averaged over all azimuthal angles of the radial ring.
  The uncertainty of $\xi_\mathrm{ord}$ is the error of the mean for the
  36 sectors; the rms variation is about 6 times larger.}
  \label{tab:rm}
  \begin{tabular}{cccccc}
  \hline
  $\Delta R$ & $RM_\mathrm{fg}$ & $RM_\mathrm{max,a}$ & $\xi_\mathrm{reg}$ & $\chi^2$ & $\xi_\mathrm{ord}$ \\
  $[$kpc$]$ & [$\radm$] & [$\radm$] & [$\degr$] &    & [$\degr$] \\
  \hline
  $7-8$   & $-128\pm7$ & $93\pm13$  & $-4\pm5$ & 2.9 & $-30\pm5$ \\
  $8-9$   & $-118\pm5$ & $99\pm10$  & $-9\pm3$ & 2.3 & $-29\pm4$ \\
  $9-10$  & $-121\pm5$ & $120\pm9$  & $-7\pm3$ & 2.4 & $-26\pm3$ \\
  $10-11$ & $-130\pm4$ & $123\pm8$  & $-7\pm2$ & 1.1 & $-27\pm2$ \\
  $11-12$ & $-130\pm6$ & $129\pm11$ & $-5\pm3$ & 1.8 & $-27\pm3$ \\
  \hline
  \end{tabular}
\end{center}
\end{table*}
\normalsize
%---------------------------------------------------------------------------
Faraday rotation is a tool used to study the pattern of the large-scale
regular field, but it is insensitive to a large-scale pattern of the anisotropic
turbulent field.
We computed Faraday rotation measures $RM = \Delta \chi /(\lambda_1^2 -\lambda_2^2)$
(where $\lambda$ is measured in metres) from the polarization
angles $\chi$ between $\lambda_1=0.1133$\,m and $\lambda_2=0.0618$\,m at
$5\arcmin$ resolution, and also between $\lambda_1=0.0618$\,m and
$\lambda_2=0.0359$\,m at $3\arcmin$ resolution (Figures~\ref{fig:rm6_11}
and \ref{fig:rm3_6}). In both figures $RM$ varies smoothly along the ring,
with the lowest values near the north-eastern major axis and the highest
near the south-western major axis. A few regions outside of the ring
(near the left edge of Fig.~\ref{fig:rm6_11}) show significant deviations from
this behaviour.
%and could be caused by $\HI$ gas in the dust clouds seen in Fig.~\ref{fig:cm11ir}.

%%----------------------------------------
%FIG25 Azimuthal variation of residuals of RM(3/6cm)
\begin{figure}[htbp]
\begin{center}
\includegraphics[width=0.6\columnwidth,angle=270]{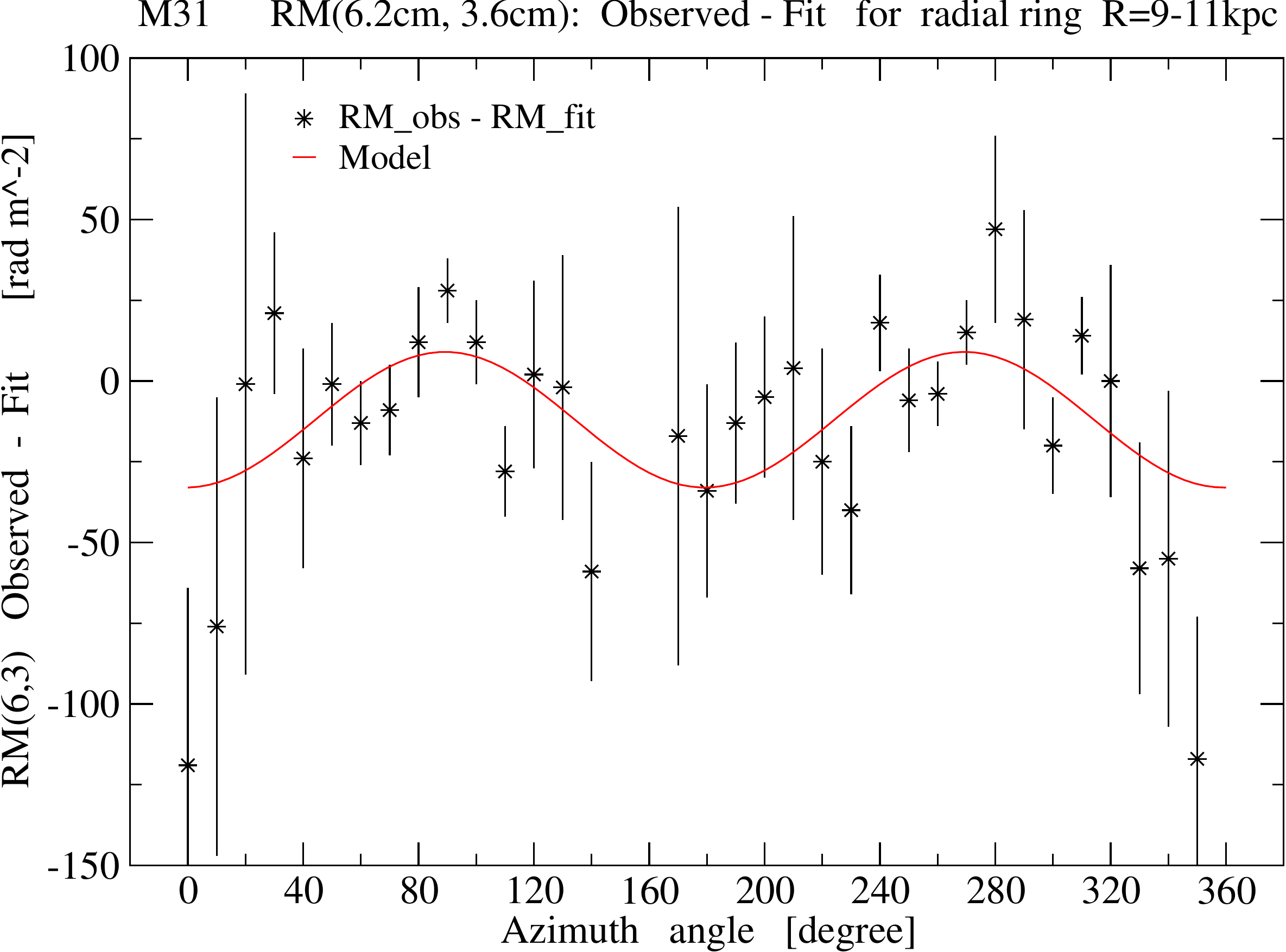}
\hfill
\caption{Variation (with azimuthal angle in the galaxy plane) of the residual
Faraday rotation measures $RM$ between \wave{6.2} and \wave{3.6} at $3\arcmin$
HPBW, averaged in the radial ring between 9\,kpc and 11\,kpc, after
subtracting the best-fitting sinusoidal model from
Fig.~\ref{fig:rm3_6_azm} (black line). The reduced $\chi^2$ value is 1.2.
%The azimuthal angle is counted from the north-eastern major axis (left side in Fig.~\ref{fig:rm3_6}).
}
\label{fig:rm3_6_azm_res}
\end{center}
\end{figure}
%%----------------------------------------

The $RMs$ in the central region are different in the two figures:  about
$-150\,\radm$ in Fig.~\ref{fig:rm6_11} and about $-100\,\radm$ in
Fig.~\ref{fig:rm3_6}. \citet{giessuebel14} measured the $RM$ between \wave{6.2}
and \wave{3.5} in the central region with $15\arcsec$ resolution and found
periodic variations of about $\pm100\,\radm$
around the foreground $RM_\mathrm{fg}\simeq-100\,\radm$, which indicates
a separate $ASS$ field in the central region. The beam width of our new
observations is too large to resolve this inner field.

If the large-scale $ASS$ pattern of the ordered field found in
Sect.~\ref{sec:PI} is also valid for the regular field (which is part of
the ordered field), the $RM$ is expected to vary as \citep[see][]{krause89a}
\begin{equation}
\label{eq:RM}
RM = RM_\mathrm{fg} + RM_\mathrm{max,a} \, cos\,\beta_{\mathrm{reg}} = RM_\mathrm{fg} + RM_\mathrm{max,a} \, cos\,(\phi - \xi_\mathrm{reg}) ~ sin\,i \, ,
\end{equation}
where $RM_\mathrm{fg}$ is the $RM$ contribution from the Milky Way in the
foreground of M~31; $\phi$ is the azimuthal angle in the galaxy plane;
$\beta_{\mathrm{reg}}$ is the aspect angle between the regular field and the
line of sight;  $\xi_\mathrm{reg}$ is the pitch angle of the regular field,
assumed to be constant along $\phi$; and $RM_\mathrm{max,a}$ is the maximum $RM$ of
the $ASS$ mode near the south-western major axis of the ring
(i.e. at the azimuthal angle $\phi=180\degr+\xi_\mathrm{reg}$).
Figures~\ref{fig:rm6_11_azm} and \ref{fig:rm3_6_azm} show that sinusoidal variations
give good fits to the data averaged in sectors of the radial ring
between 9\,kpc and 11\,kpc.

The foreground $RMs$ of the sinusoidal fits in the radial ring 9--11\,kpc
are similar, $RM_\mathrm{fg}=(-118\pm3)\,\radm$ between \wave{11.3} and \wave{6.2}
(Fig.~\ref{fig:rm6_11_azm}) and $RM_\mathrm{fg}=(-125\pm4)\,\radm$ between
\wave{6.2} and \wave{3.6} (Fig.~\ref{fig:rm3_6_azm}, black curve).
The amplitude $RM_\mathrm{max,a}=(78\pm6)\,\radm$ between \wave{11.3} and
\wave{6.2} is lower compared to $RM_\mathrm{max,a}=(118\pm9)\,\radm$
between \wave{6.2} and \wave{3.6} in the same radial range by a factor of
$\simeq1.5$. As this factor characterizes the relative amount of Faraday
depolarization, M~31 is less transparent to polarized emission
(partly `Faraday thick') at \wave{11.3} compared to \wave{6.2} and \wave{3.6}.

We performed sinusoidal fits according to Eq.~\ref{eq:RM} for the $RM$ data
between \wave{6.2} and \wave{3.6} for five radial rings between 7\,kpc and
12\,kpc {in the galaxy plane.
The results are given in Table~\ref{tab:rm}. The average
$RM_\mathrm{fg}$ of $(-125\pm5)\,\radm$ is smaller than the values
of about $-90\,\radm$ found based on previous data \citep{beck82,berkhuijsen03},
but the new data are more reliable. The value of
$RM_\mathrm{fg}$ does not show a significant variation with radius, indicating that
$RMs$ from the Milky Way do not vary on an angular scale similar to
that of the ring width in M~31.
The amplitudes $RM_\mathrm{max,a}$ increase with radius, indicating that
the strength of the regular field of M~31 increases outwards, as already
noted by \citet{fletcher04}.
The pitch angles $\xi_\mathrm{reg}$ are similar to the pitch angle of the
gaseous spiral arms of about $-7\degr$ \citep{arp64,braun91,chemin09}.

Linear models of $\alpha$--$\Omega$ dynamo action in galaxies \citep{shukurov00}
predict that the absolute value of the pitch angle $|\xi_\mathrm{reg}|$
is constant for a
flat rotation curve, but that it decreases with increasing radius if the scale height of
the gas disc increases (`flaring disc'). According to the non-linear dynamo
model developed by \citet{chamandy15b}, the magnetic pitch angle, in the saturated state of
field evolution, depends on several parameters that may vary differently with radius.
Our results (Table~\ref{tab:rm}) indicate that $\xi_\mathrm{reg}$ is about
constant with radius, consistent with the prediction from the simple model.
On the other hand, the mode analysis of multi-frequency polarization angles
used by \citet{fletcher04} yielded larger values of $\xi_\mathrm{reg}$ between $-11\degr$
and $-19\degr$ in  radial rings similar to those in Table~\ref{tab:rm}
and a hint of a radial decrease. However, anisotropic turbulent fields (which affect
polarization angles and intensities but not $RMs$) are neglected in the method of
\citet{fletcher04}, so that their values of $\xi_\mathrm{reg}$ are correct
only if $\xi_\mathrm{ord} \simeq \xi_\mathrm{reg}$.

The azimuthal $RM$ variation in Fig.~\ref{fig:rm3_6_azm} shows significant
deviations from the sinusoidal fit (black curve). \citet{sofue87a} suggested
the existence of a  $BSS$ mode superimposed onto the $ASS$ mode.
The azimuthal variation of $RM$ for a $BSS$ field is \citep{krause90}

\begin{equation}
\label{eq:RM2}
RM = RM_\mathrm{fg} + RM_\mathrm{max,b} \, cos\,(2\,\phi - \delta) ~ sin\,i \, ,
\end{equation}
where $\phi$ is the azimuthal angle in the galaxy plane and $\delta$
the phase, which is related to the pitch angle and the position angle of the spiral
pattern in the galaxy plane.

The red curve in Fig.~\ref{fig:rm3_6_azm} shows the fit for the combined
$ASS + BSS$ field, for the radial range $9-11$\,kpc where the
signal-to-noise ratios are highest. The amplitude of the $BSS$ field
is $RM_\mathrm{max,b}=(21\pm7)\radm$, about six times smaller than the amplitude
$RM_\mathrm{max,a}$ of the $ASS$ mode. The fit of the combined modes is not
statistically better because the $BSS$ field is weak. We computed
the residuals from the $ASS$ fit (Fig.~\ref{fig:rm3_6_azm_res}), which  vary double-periodically with azimuthal angle and hence support
the superimposed $BSS$ field.

%%----------------------------------------
%FIG26 Azimuthal variation of PW(3/6cm)
\begin{figure}[htbp]
\begin{center}
\includegraphics[width=0.6\columnwidth,angle=270]{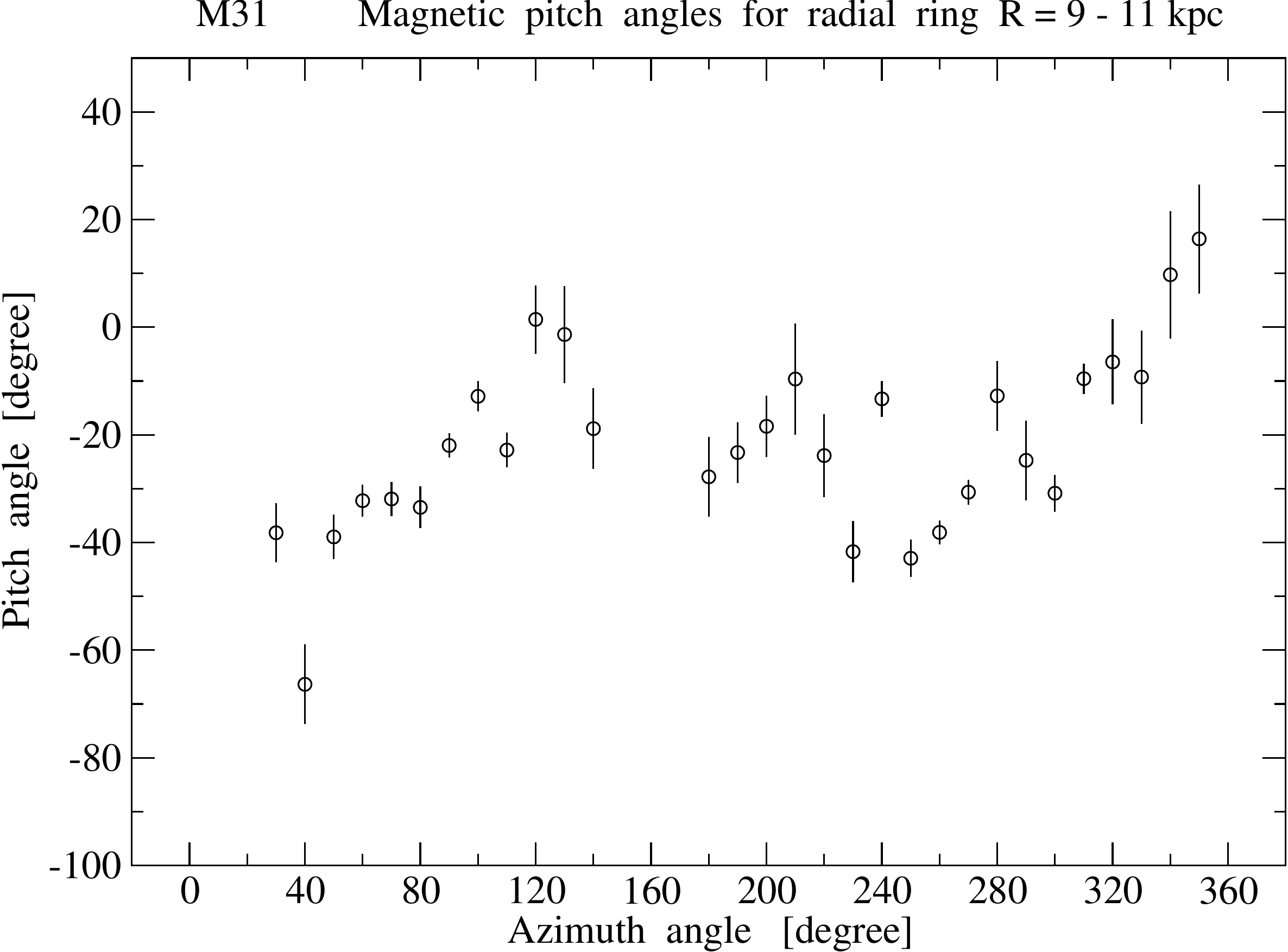}
\hfill
\caption{Variation (with azimuthal angle in the galaxy plane) of the
  intrinsic pitch angle $\xi_\mathrm{ord}$ of the ordered magnetic field,
  calculated from the intrinsic orientations of the ordered field at
  $3\arcmin$ HPBW, corrected with $RMs$
  between \wave{6.2} and \wave{3.6}, and averaged in the radial ring between
  9\,kpc and 11\,kpc.}
\label{fig:pw3_6_azm}
\end{center}
\end{figure}
%%----------------------------------------

Faraday rotation allows us to compute the intrinsic pitch angle
$\xi_\mathrm{ord}$ of the ordered field from the observed orientation
of the polarized emission in two steps, to be compared with $\xi_\mathrm{reg}$
as computed above.
Firstly, the $RMs$ shown in Figs.~\ref{fig:rm6_11} or \ref{fig:rm3_6}
were used to correct the orientation $\chi_\mathrm{obs}$ of the ordered magnetic
field in the plane of the sky observed at wavelength $\lambda$ for Faraday rotation
in order to achieve the intrinsic orientation (i.e. at infinitely small
wavelength) in the plane of the sky via
$\chi_\mathrm{ord}= \chi_\mathrm{obs} - RM ~ \lambda^2$.
Then the intrinsic pitch angle $\xi_\mathrm{ord}$
of the ordered field in the galaxy plane follows from
\begin{equation}
\label{eq:pitch}
\xi_\mathrm{ord} ~ = ~ \phi + 90\degr - arctan \, [\, tan(\chi_\mathrm{ord}-\chi_\mathrm{p}) \, / \, cos\,i \,] \, ,
\end{equation}
where $\chi_\mathrm{ord}$ is the intrinsic position angle of the ordered
field in the plane of the sky and $\chi_\mathrm{p}$ is the position angle of the
major axis of the galaxy in the plane of the sky.

The azimuthal variation of the intrinsic pitch angle $\xi_\mathrm{ord}$
in the ring 9--11\,kpc, calculated from the intrinsic orientations of the
ordered field at $3\arcmin$ resolution, corrected with $RMs$
between \wave{6.2} and \wave{3.6}, is shown in
Figure~\ref{fig:pw3_6_azm}. The average pitch angle is $-26\degr \pm3 \degr$
(weighted according to error bars). The average pitch angle derived from
the intrinsic polarization angles at $5\arcmin$ resolution, corrected with
$RMs$ between \wave{11.3} and \wave{6.2}, in the same
radial range 9--11\,kpc is $-17\degr \pm6 \degr$ (also weighted according
to error bars), consistent with the value obtained using the smaller wavelengths.
In order to search for any radial variation, we computed averages of
$\xi_\mathrm{ord}$ in five radial rings, given in the last column of
Table~\ref{tab:rm}. No significant variation with radius is found.

The large variations in the intrinsic pitch angles $\xi_\mathrm{ord}$
seen in Fig.~\ref{fig:pw3_6_azm} indicate that the field structure is more
complex than an $ASS + BSS$ field.
%The weak $BSS$ mode
%(Fig.~\ref{fig:rm3_6_azm_res}) does not significantly affect the pitch angles.
The value of $\xi_\mathrm{ord}$ jumps by about $70\degr$ from positive to
negative values near the north-eastern major axis ($\phi \approx 0\degr$), and
$\xi_\mathrm{ord}$ calculated from the intrinsic orientations of the ordered
field, corrected with $RMs$ between \wave{11.3} and \wave{6.2},
shows a similar behaviour.
%This could be related to the deviation in polarized
%intensity detected in the 4th quadrant (Fig.~\ref{fig:PI_quadrants}).
We propose that $\xi_\mathrm{ord}$ is affected by local field
deviations and/or anisotropic turbulent fields.

The averages of $\xi_\mathrm{ord}$ in Table~\ref{tab:rm} are significantly
smaller than those of $\xi_\mathrm{reg}$ and the pitch angle of the gaseous
spiral arms of about $-7\degr$.
Pitch angles of the ordered field that deviate from those of the gaseous spiral
arms have also been found in the spiral galaxies M~74 \citep{mulcahy17},
M~83 \citep{frick16}, and M~101 \citep{berkhuijsen16}. These results were
believed to show that the mean-field dynamo does not generate regular
fields that are aligned with the spiral arms because $\xi_\mathrm{reg}$ depends on
several parameters that are unrelated to spiral arms \citep{chamandy15b}.

The results for M~31 presented here suggest a different interpretation:
$B_\mathrm{ord}$ has two components, the regular field $B_\mathrm{reg}$
and the anisotropic turbulent field $B_\mathrm{an}$. As $\xi_\mathrm{reg}$
in Table~\ref{tab:rm} is similar to the pitch angle of the gaseous spiral
arms of about $-7\degr$, the deviation may arise in the anisotropic
turbulent field. The regular field $B_\mathrm{reg}$ and the anisotropic
turbulent field $B_\mathrm{an}$ have different spiral patterns that may be
shaped by different physical processes.

% {\bf Rainer, there is a tendency of high pitch angles where TH emission
%  is low. An influence of TH on pitch angles could perhaps mean that these
%  pitch angles are not for a pure regular field, but that some turbulent
%  field is still contained in this regular field. What do you think?
%  --  Maybe an effect of Faraday depolarization?}

%----------------------------SECTION-----8-----------------------------

\section{Summary and conclusions}
\label{sec:conclusions}

In order to study the magnetic field structure of M~31, we used the
Effelsberg telescope to perform three new deep radio continuum surveys
in total intensity and polarization at the wavelengths of \wave{11.3}
(2.645\,GHz), \wave{6.2} (4.85\,GHz), and \wave{3.6} (8.35\,GHz).
The angular resolutions (HPBW) are $4\farcm4$, $2\farcm6$, and $1\farcm6$,
respectively (Table~\ref{tab:surveys}).
As we wanted to study the large-scale emission, we subtracted point
sources unrelated to M~31 at each wavelength. The resulting maps are
shown in Figure~\ref{fig:cm11i} to Figure~\ref{fig:cm3pi}.

In this first paper we have presented the observations and reduction procedures and
discussed results on the spectral index and radial scale lengths, and on  Faraday rotation measures and pitch angles of the regular magnetic
field. Below we summarize our main conclusions:

1. At all wavelengths the well-known emission ring between about 7\,kpc and
13\,kpc radius in the galaxy plane
stands out, both in I and in PI. The nuclear region
is also very bright. PI is low on the major axis of the ring
in the plane of the sky on both sides, which is a
first indication that the ordered magnetic field is oriented along the bright
ring, so that the field component in the plane of the sky is smallest
near the major axis.

2. Including all available surveys between 0.3\,GHz and 4.85\,GHz,
covering M~31 to at least $R = 16$\,kpc, we find a spectral index of
the integrated total emission of $\alpha = 0.71\pm0.02$
(defined as $S \propto \nu^{-\alpha}$). After
subtraction of the thermal emission, we obtain a spectral index of
the integrated non-thermal emission of $\alphan = 0.81\pm0.03$.

3. The spectral indices vary across M~31. Maps of total and non-thermal
spectral indices between \wave{21.1} and \wave{3.6} at $1\farcm5$
resolution show that $\alpha$ ($\alphan$) is about 0.4 (0.5) near
star-forming regions and steepens to about 1.0 towards the inner and
outer parts of the bright ring due to radiation losses of the CREs
propagating away from their birth places near the star-forming regions.

4. The radial variation of the intensities of I, NTH, PI, and TH all
peak between 8 and 12\,kpc radius, but their scale lengths ($L$) differ.
At \wave{6.2} and $3\arcmin$ resolution, between $R = 11$\,kpc and
$R = 15$\,kpc, we find  $L(I) = (3.02\pm0.11)$\,kpc,
$L(NTH) = (3.66\pm0.10)$\,kpc,
$L(PI) = (4.84\pm0.16)$\,kpc, and $L(TH) = (1.87\pm0.05)$\,kpc. The
longer scale lengths of NTH and PI than of TH again indicate energy
losses of the CREs when travelling away from their birthplaces near star-forming
regions. The difference in scale lengths shows that the propagation length
of the CREs could be several kpc.

5. The polarized intensity, averaged in azimuthal sectors in
the radial ring between 9\,kpc and 11\,kpc in the galaxy plane,
varies with azimuthal angle as a double-periodic
curve, with maxima near the minor axis and minima near the major axis.
%The fits give phase shifts of $-9\degr \pm 3\degr$ at \wave{3.6} and $-14\degr \pm 2\degr$ at \wave{6.2}.
This indicates that the ordered magnetic field is  oriented almost along
the ring but has a small pitch angle.

6. Faraday rotation measures ($RMs$) between \wave{11.3} and \wave{6.2}
and between \wave{6.2} and \wave{3.6}, averaged in
azimuthal sectors of the radial ring between 9\,kpc and 11\,kpc,
vary smoothly along the emission ring as a cosine function with azimuthal
angle in the galaxy plane, which is a signature of a regular field with
an axisymmetric spiral ($ASS$) pattern. The phase shift of the variation
of $-7 \degr \pm2 \degr$ is interpreted as the average spiral
pitch angle of the regular field. It shows no significant variation
with radius and is similar to the pitch angle of the gaseous spiral arms.

7. The residuals between the $RMs$ computed between \wave{6.2} and
\wave{3.6} and the $ASS$ fit
show a double-periodic variation with azimuthal angle, indicative
of a superimposed  $BSS$ mode of the regular field,
with an about six times smaller amplitude compared to the $ASS$ mode.

8. The amplitude of the $RMs$ between \wave{11.3} and \wave{6.2}
between 9\,kpc and 11\,kpc radius
is about 1.5 times lower than between \wave{6.2} and \wave{3.6},
indicating that Faraday depolarization at \wave{11.3} is stronger
(greater Faraday thickness) than at \wave{6.2} and \wave{3.6}).

9. The {average pitch angle of the ordered field, derived
from the intrinsic orientations of the polarized emission
between 9\,kpc and 11\,kpc radius, is $-26\degr\pm3\degr$.
The difference in pitch angles of regular and
ordered field indicates that the ordered field contains a significant
fraction of an anisotropic turbulent field that has a
pattern that is different to the regular ($ASS + BSS$) field.
\\

New insights to the magnetic field of M~31, especially measuring
the extent in the outer disc and the search for large-scale reversals,
can be expected from deep observations of polarized background sources
in L band (\wave{20}, hence improving the pioneering work performed by \citet{han98}.
While the southern location of the Square Kilometre Array (SKA, under
construction) hampers observation of M~31, the Jansky Very Large Array
(JVLA) and APERTIF \citep{oosterloo18} are suitable instruments for such
investigations.
\\

\begin{acknowledgements}
  We thank the operators at the Effelsberg telescope for support during
  on-site observations and for supervising remote observations. We thank
  Patricia Reich for her support at various stages of data processing,
  especially for writing the {\tt turboplait} script and for an early version
  of the script {\tt subtrans}. Peter M\"uller is acknowledged for making several
  data processing scripts available to us already during the development
  of {\tt NOD3}. We thank Marita Krause, Andrew Fletcher, and the anonymous
  referee for careful reading of the manuscript and useful suggestions.
%This work is based on observations with the 100-m telescope of the MPIfR (Max-Planck-Institut f\"ur Radioastronomie) at Effelsberg. --

\end{acknowledgements}


%
%%%%%%%%%%%%%%%%%%%%%%%%%%%%%%%%%%%%%%%%%%%%%%%%%%%%%%%%%%%%%%

\bibliographystyle{aa} % style aa.bst
\bibliography{m31}

\end{document}